\documentclass[fleqn,usenatbib]{mnras}

\usepackage{newtxtext,newtxmath}

\usepackage[T1]{fontenc}
\usepackage{ae,aecompl}

\usepackage{graphicx}	
\usepackage{amsmath}	
\usepackage{amssymb}	

\newcommand{\overbar}[1]{\mkern 1.5mu\overline{\mkern-1.5mu#1\mkern-1.5mu}\mkern 1.5mu}
\usepackage[flushleft]{threeparttable}

\title[ASAS-SN Catalog of Variable Stars I]{The ASAS-SN Catalog of Variable Stars I: \textit{The Serendipitous Survey}}

\author[T. Jayasinghe et al.]{T. Jayasinghe$^{1,2}$\thanks{E-mail: jayasinghearachchilage.1@osu.edu},
C. S. Kochanek$^{1,2}$,
K. Z. Stanek$^{1,2}$,
B. J. Shappee$^{3}$,
\newauthor 
T. W. -S. Holoien$^{4}$,
Todd A. Thompson$^{1,2}$,
J. L. Prieto$^{5,6}$,
Subo Dong$^{7}$,
M. Pawlak$^{8}$,
\newauthor 
J. V. Shields$^{1}$,
G. Pojmanski$^{9}$,
S. Otero$^{10}$,
C. A. Britt$^{11}$,
D. Will$^{1,11}$
\\
$^{1}$Department of Astronomy, The Ohio State University, 140 West 18th Avenue, Columbus, OH 43210, USA\\
$^{2}$Center for Cosmology and Astroparticle Physics, The Ohio State University, 191 W. Woodruff Avenue, Columbus, OH 43210, USA\\
$^{3}$Institute for Astronomy, University of Hawaii, 2680 Woodlawn Drive, Honolulu, HI 96822,USA\\
$^{4}$Carnegie Observatories, 813 Santa Barbara Street, Pasadena, CA 91101, USA\\
$^{5}$N\'ucleo de Astronom\'ia de la Facultad de Ingenier\'ia y Ciencias, Universidad Diego Portales, Av. Ej\'ercito 441, Santiago, Chile\\
$^{6}$Millennium Institute of Astrophysics, Santiago, Chile\\
$^{7}$Kavli Institute for Astronomy and Astrophysics, Peking University, Yi He Yuan Road 5, Hai Dian District, China\\
$^{8}$Institute of Theoretical Physics, Faculty of Mathematics and Physics, Charles University in Prague, Czech Republic\\
$^{9}$Warsaw University Observatory, Al Ujazdowskie 4, 00-478 Warsaw, Poland\\
$^{10}$The American Association of Variable Star Observers, 49 Bay State Road, Cambridge, MA 02138, USA\\
$^{11}$ASC Technology Services, 433 Mendenhall Laboratory 125 South Oval Mall Columbus OH, 43210, USA\\
}

\date{Accepted XXX. Received YYY; in original form ZZZ}

\pubyear{2018}

\begin{document}
\label{firstpage}
\pagerange{\pageref{firstpage}--\pageref{lastpage}}
\maketitle

\begin{abstract}
The All-Sky Automated Survey for Supernovae (ASAS-SN) is the first optical survey to routinely monitor the whole sky with a cadence of $\sim2-3$ days down to V$\lesssim17$ mag. ASAS-SN has monitored the whole sky since 2014, collecting $\sim100-500$ epochs of observations per field. The V-band light curves for candidate variables identified during the search for supernovae are classified using a random forest classifier and visually verified. We present a catalog of 66,179 bright, new variable stars discovered during our search for supernovae, including 27,479 periodic variables and 38,700 irregular variables. V-band light curves for the ASAS-SN variables are available through the ASAS-SN variable stars database
(https://asas-sn.osu.edu/variables). The database will begin to include the light curves of known variable stars in the near future along with the results for a systematic, all-sky variability survey.
\end{abstract}




\section{Introduction}
The dream of observing the entire sky all the time is not new---indeed, already in 1906, Pickering ``wanted to test the constancy" of the brightest fifty million stars (see \citealt{sobel}). That desire was clearly motivated by the excellent results obtained by the Harvard variable stars research group (e.g., \citealt{1904HarCi..90....1L} ; \citealt{1908AnHar..60...87L}; \citealt{1912HarCi.173....1L} ; \citealt{1938AnAp....1..430P}), which for many years dominated this field of study. \citet{2000PASP..112.1281P} discussed the importance of all-sky variability surveys and advocated for the research that could be carried out with small telescopes, a principle out of which the All-Sky Automated Survey (ASAS; \citealt{2002AcA....52..397P}) and the All-Sky Automated Survey for SuperNovae (ASAS-SN, \citealt{2014ApJ...788...48S, 2017PASP..129j4502K}) originated.

ASAS-SN is the first ground-based survey to monitor the entire visible sky to a depth of $V\lesssim17$ mag on a regular basis. ASAS shared a similar goal at $V\lesssim14$ mag, and data from the ASAS project for declinations $\delta<28$ deg \citep{2002AcA....52..397P,2003AcA....53..341P,2004AcA....54..153P,2005AcA....55...97P,2005AcA....55..275P} have been used extensively. While the current ASAS-SN data was obtained on a cadence of 2-3 days, ASAS-SN has recently added 3 new ASAS-SN units (each consisting of four telescopes) at CTIO, Chile (``Bohdan Paczynski"), McDonald Observatory, Texas (``Henrietta Leavitt") and the South African Astrophysical Observatory (SAAO), South Africa (``Cecilia Payne-Gaposchkin"), which has significantly improved cadence. These new units use the SDSS g-band filter. The ASAS-SN telescopes are hosted by the Las Cumbres Observatory (LCO; \citealt{2013PASP..125.1031B}).

Modern surveys such as ASAS \citep{2002AcA....52..397P}, the Optical Gravitational Lensing Experiment (OGLE; \citealt{2003AcA....53..291U}), the Northern Sky Variability Survey (NSVS; \citealt{2004AJ....127.2436W}), MACHO \citep{1997ApJ...486..697A}, EROS \citep{2002A&A...389..149D}, Pan-STARRS1 \citep{2016arXiv161205560C}, the Palomar Transient Factory (PTF; \citealt{2009PASP..121.1395L}) and the Catalina Real-Time Transient Survey (CRTS; \citealt{2014ApJS..213....9D}) have been very productive and have collectively discovered $\sim 10^6$ variables. The study of variable stars has led to many key astrophysical insights governing stellar theory, multiplicity and cosmology (see \citealt{1963ARA&A...1..367Z,1978ARA&A..16....1P,2010A&ARv..18...67T}, and references therein).

ASAS-SN has focused on the detection of bright supernovae with minimal observational bias (e.g., \citealt{2017MNRAS.471.4966H}). However, many other transient and variable sources have been discovered during this process. These include roughly $\sim$90,000 candidate variable sources that were flagged during the search for supernovae. In this work, we detail the steps taken to classify these variables and create a catalog of 66,179 previously undiscovered variables. In Section \ref{data}, we discuss the ASAS-SN observations and data reduction leading to the production of light curves for the variable sources. Section \ref{varclass} presents the variability classification procedure based on random forest classification models. We discuss the catalog and our results in Section \ref{catalog} and present a summary of our work in Section \ref{conclude}.

\section{Observations and Data reduction}
\label{data}
This catalog was compiled using observations taken by the ``Brutus" (Haleakala, Hawaii) and ``Cassius" (CTIO, Chile) quadruple telescopes between 2013 and 2017 with $\sim$ 100-500 epochs of observation for each field to a depth of $V\lesssim17$ mag. The field of view of a single ASAS-SN camera is 4.5 deg$^2$, the pixel scale is 8\farcs0 and the FWHM is $\sim$ 2 pixels. The ASAS-SN cameras capture three 90 sec images for every epoch which are merged to improve the signal-to-noise. ASAS-SN saturates at $\sim 10-11$ mag depending on the camera and image position \citep{2017PASP..129j4502K}.

ASAS-SN is primarily a transient search and makes use of the ISIS image subtraction pipeline \citep{1998ApJ...503..325A,2000A&AS..144..363A} to process images. For the variable light curves, we performed aperture photometry on the difference images using the IRAF \verb apphot \space package and calibrated the results using the AAVSO Photometric All-Sky Survey (APASS; \citealt{2015AAS...22533616H}). We extracted light curves for the variable sources using a 2 pixel radius aperture. We rejected images taken in poor weather conditions, images that were out of focus with FWHM $>2$ pixels, images that had poor astrometric solutions, and images where the variable source is within 0.2 deg of a detector edge. 

Since ASAS-SN first began its search for supernovae in 2013 \citep{2014ApJ...788...48S}, previously undiscovered variable objects were flagged as candidates for variable stars. This list was cleaned of sources with bad photometry, inadequately sampled light curves ($N<30$ epochs) and inaccurate astrometry, ultimately resulting in the list of $\sim 80,000$ potentially variable sources we analyze here.

The sky is divided into fields corresponding to the field of view of a camera. The 4 cameras in a single mount have overlapping fields of view, and equatorial fields can be observed by both units. Thus, over $\sim 45\%$ of the variable sources have photometry from two or more cameras. Calibrated light curves are extracted for each camera independently, and there can be small calibration offsets between them. For each source, the camera with the largest number of measurements is considered the primary. We consider other cameras only if they have made at least 50 measurements, to ensure that sufficient phase coverage is available for periodic variables. The photometry from each secondary camera is offset by the differences between the light curve medians for the secondary camera and the primary camera. This method works well for low amplitude variables, but can work poorly compared to the photometric uncertainties for high amplitude, long period variables. 

\begin{figure*}
	\includegraphics[width=\textwidth]{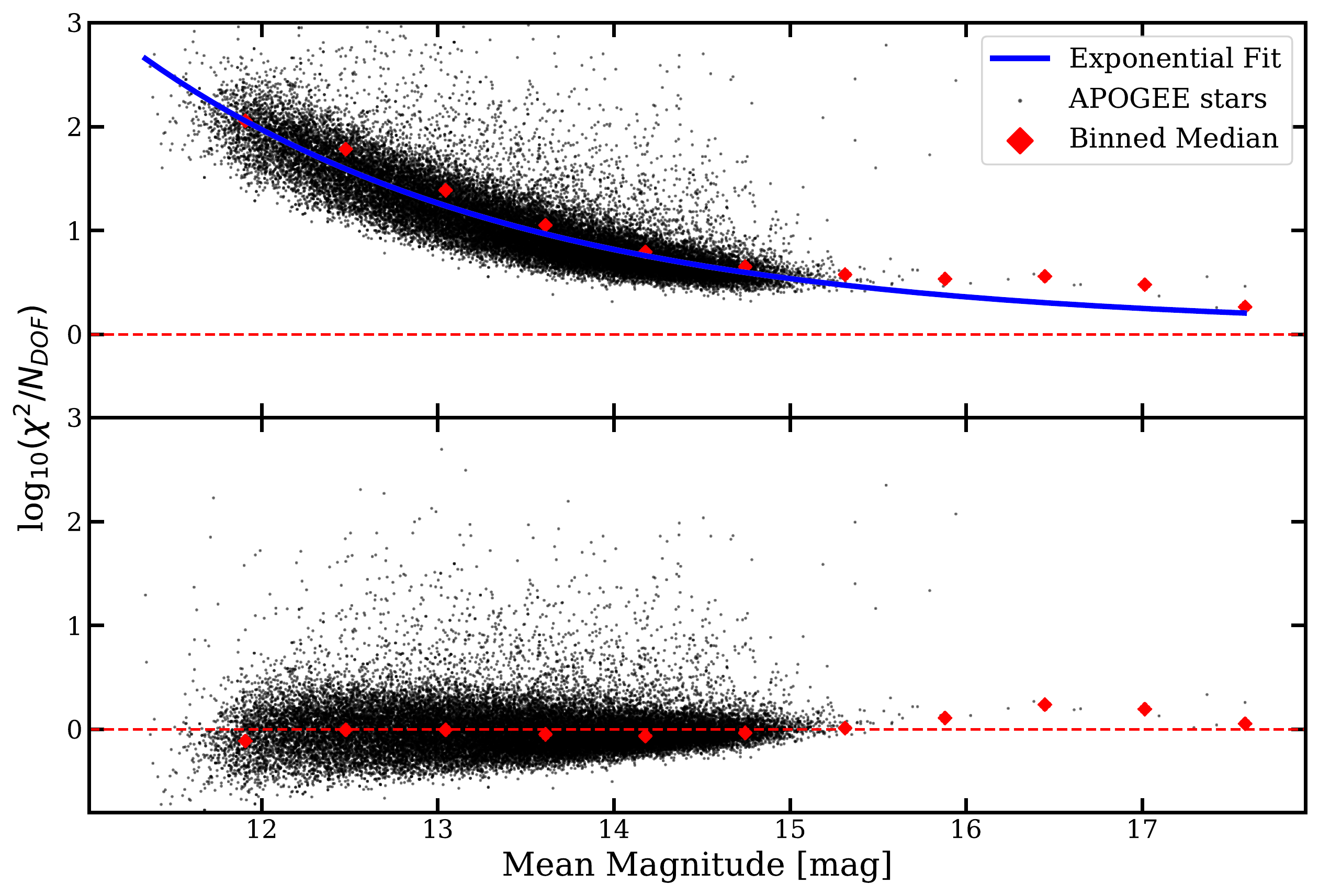}
    \caption{The top panel shows the raw distribution of $\log(\chi^2/N_{DOF})$ for the APOGEE stars prior to the rescaling process. The red dashed line illustrates the expected value of $\chi^2/N_{DOF}=1$. The red points are the binned median values of $\log(\chi^2/N_{DOF})$ for this sample in bins of $\sim0.5$ mag. The curve shows an exponential model (Equation \ref{eq:errfix}) of the trend and the bottom panels shows the distribution of $\chi^2/N_{DOF}$ after rescaling the errors.}
    \label{fig:fig1}
\end{figure*}

Uncertainties estimated by local photon counting statistics tend to underestimate the light curve uncertainties. We use a procedure similar to \citet{2004AJ....128.1761H} to rescale these uncertainty estimates. We used $\sim 35,000$ sources from the APOGEE \citep{2015AJ....150..148H} survey for this process. The reduced $\chi^2$ statistic, \begin{equation}
    \frac{\chi^2}{N_{DOF}}=\frac{1}{N-1}\sum_{i=1}^{N} \left(\frac{V_i -\overbar{V}}{\sigma_i}\right)^2\,,
	\label{eq:chi2}
\end{equation}is calculated for each source, where $V_i$ is the V-band magnitude for epoch $i$ , $\overbar{V}$ is the mean V-band magnitude and $\sigma_i$ is the formal error of the magnitude measurement. Figure \ref{fig:fig1} shows the ${\chi^2}/{N_{DOF}}$ distribution for the APOGEE stars. We find the usual pattern that errors are underestimated and that the fractional underestimates are larger for brighter stars. We fit an exponential, \begin{equation}
	f(\overbar{V})=Ae^{-B\overbar{V}}+C\,,
	\label{eq:errfix}
\end{equation} with $A=4.98\times 10^2$, $B=4.64\times 10^{-1}$ and $C=6.31 \times 10^{-2}$ to the observed $\log(\chi^2/N_{dof})$ versus $\overbar{V}$ distribution. To correct the errors in the light curves, we multiply the formal magnitude errors by $(10^{f(\overbar{V})})^{\frac{1}{2}}$. The results of this correction are illustrated in the bottom panel in Figure \ref{fig:fig1}.

\section{Variability Classification}
\label{varclass}
Here we describe the procedure we took to classify the variable sources. In Section $\S3.1$, we describe the procedure we took to cross match our list of variables against catalogs of discovered variables. We describe our approach to derive periods for these sources in Section $\S3.2$ and the first step we took to classify these variables in Section $\S3.3$. We discuss our final random forest classifier model in Section $\S3.4$.

\subsection{Crossmatches to catalogs}
Despite the relatively large pixel scale of $\sim$8", ASAS-SN's typical astrometric error is $\sim$1". We update the coordinates of the variables detected in the ASAS-SN pipeline to those of a 2MASS \citep{2006AJ....131.1163S}, UCAC4 \citep{2013AJ....145...44Z}, NOMAD \citep{2004AAS...205.4815Z},  USNOB1.0 \citep{2003AJ....125..984M} or PPMXL \citep{2010AJ....139.2440R} star of the correct magnitude within 10\farcs0 of the ASAS-SN coordinate. This also provides near-infrared and optical colors for use in classification. We find that 99\% of the new variables are within 5\farcs2 of such a cataloged star.

We matched our list of variables to the VSX \citep{2006SASS...25...47W} and GCVS \citep{2017ARep...61...80S} catalogs available in October 2017, with a matching radius of 16\farcs0, to identify previously discovered variables. The VSX catalog is likely the most complete catalog of known variables in the modern era \citep{2006SASS...25...47W}, and includes variables discovered by a large number of surveys. At the time of the matching the VSX catalog contained $\sim 460,000$ variables. We also match our variables to the recent OGLE-IV catalogs of eclipsing binaries in the Magellanic clouds \citep{2016AcA....66..421P} and the Galactic bulge \citep{2016AcA....66..405S}, along with the OGLE-IV catalog of Cepheids in the Magellanic clouds \citep{2015AcA....65..297S} as these have not yet been merged with the VSX catalog.
 
\subsection{Period Determination}
We used the Generalized Lombe-Scargle (GLS, \citealt{2009A&A...496..577Z,1982ApJ...263..835S}), the Multi-Harmonic Analysis Of Variance (MHAOV, \citealt{1996ApJ...460L.107S}), the Phase Dispersion Minimization (PDM, \citealt{1978ApJ...224..953S}) and the Box Least Squares (BLS, \citealt{2002A&A...391..369K}) periodograms to search for periodicity. We use the \verb"astropy" implementation of the GLS algorithm \citep{2013A&A...558A..33A}, the \verb"P4J" implementation of the MHAOV periodogram \citep{2017arXiv170903541H}, the \verb"pwkit" implementation of the PDM periodogram \citep{2017ascl.soft04001W}, and the \verb"astrobase"  implementation of the BLS periodogram \citep{astrob}.

To minimize the effects of outliers when searching for periods, we only use the epochs with magnitudes between the 5$^{th}$ and 95$^{th}$ percentiles as input to the periodograms. We searched for periods in the range $0.05\leq P \leq1000$ days. The MHAOV periodogram was initialized with $N_{harm}=2$ harmonics. The BLS periodogram was initialized with 200 phase bins and a minimum (maximum) transit duration of $0.1$ $(0.3)$ in phase. For the GLS and MHAOV periodograms, periods were selected with SNR $>5$, while BLS periods were selected with the BLS power $<0.3$. Sources that do not have a single period based on these criteria are flagged as candidate irregular variables. The 5 best periods from each periodogram were passed to the PDM algorithm as test periods to minimize the dispersion of the light curve in 20 phase bins. The overall best period for each variable was defined to be the best PDM period.  

During the first stage of our variability classification process, we utilized just the GLS, MHAOV and PDM periodograms. From the visual review of these classifications, we noted that our pipeline was unable to retrieve periods for a large fraction of the detached (Algol-type) eclipsing binaries. This can be attributed to the non-sinusoidal shapes of the eclipses seen in these light curves. We used the BLS algorithm in the next stage of our classification process to solve this problem. Unlike the GLS and MHAOV algorithms, the BLS algorithm models a transit when searching for periods and is successfully able to derive periods for detached eclipsing binaries. Furthermore, Mira-like variability was best described by GLS periods, so we only used the GLS periodogram for variable sources with amplitudes $A> 2$ mag to minimize computational time.

In the case of eclipsing binaries, the periods returned by the periodograms are frequently a sub-harmonic of the true period, usually with $P=P_{\rm true}/2$. \citet{2016MNRAS.456.2260A} describes a possible remedy to this problem by comparing the difference between minima in phase folded K2 lightcurves. Periods are doubled when the minima are different by a user-defined constant. In the case of W UMa type contact binaries, the minima in the phase folded curve can be very similar, which is problematic for this method. \citet{2013MNRAS.434.3423G} suggests that the best approach to deal with the periods for eclipsing binaries is to simply double the period once it is confirmed that the light curve is that of an eclipsing binary. 

\begin{figure*}
	\includegraphics[width=1.1\textwidth]{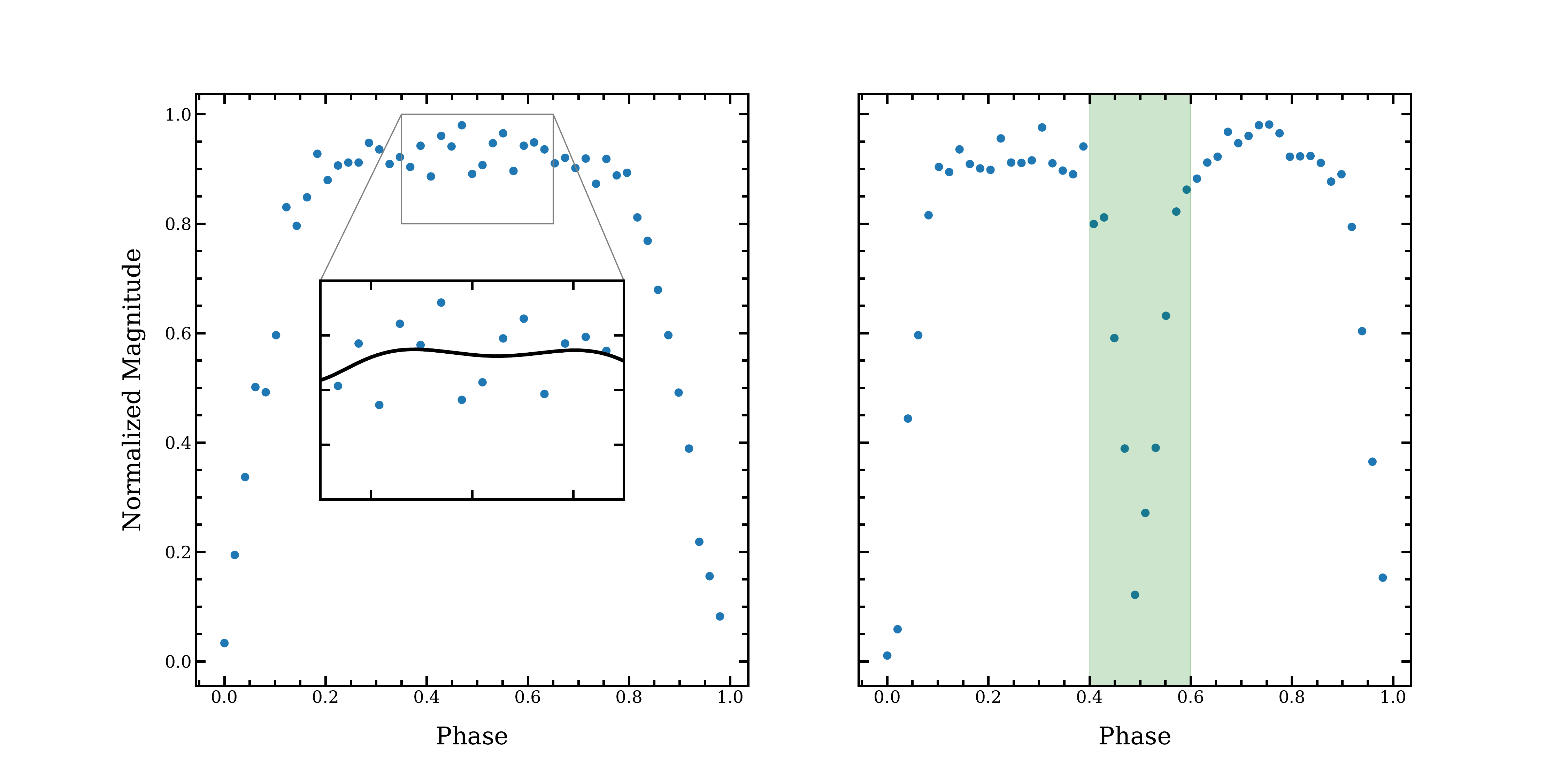}
    \caption{An illustration of the automated period doubling routine used on the ASAS-SN eclipsing binaries. \textit{Left:} The original binned ($N_{bin}=50$), phase folded light curve for the detached eclipsing binary ASASSN-V J203611.47+003921.1 with $P=0.25357$ d. The inset shows the region in phase space where the secondary minimum is expected to occur. The 6$^{th}$ order polynomial fit used in the analysis is shown in black. \textit{Right:} The final binned ($N_{bin}=50$), phase folded light curve with $P=0.50713$ d. Note the emergence of a secondary minimum of different depth after the period is doubled.}
    \label{fig:fig2}
\end{figure*}

We utilize a similar approach to that of \citet{2016MNRAS.456.2260A}. Once a variable is classified as a candidate eclipsing binary, the normalized light curve is divided into 20 bins, phase folded by the best PDM period with the primary minimum at phase 0 and normalized by the global minimum and maximum. To search for the secondary minimum, we select all the points with phases in the interval $[0.3,0.7]$, and fit a 6$^{th}$ order polynomial to this data. If the global minimum of the polynomial is between phases 0.4 and 0.6, we consider the candidate period to be the true period. If the global minimum for the polynomial is beyond this range in phase space, we double the period. We find this method to be fairly robust, although it does fail in a limited number of cases where the light curves are noisy. This method could also fail for the light curves of eccentric binaries,  and when the depth of the secondary minimum is comparable to the noise in the light curve. Figure \ref{fig:fig2} illustrates this automated period doubling routine for the eclipsing binary candidate ASASSN-V J203611.47+003921.1.

\subsection{Using the \textit{Upsilon} classifier} \label{upsclass}
For the first stage of our classification process, we used the open-source random forest classifier \textit{Upsilon} \citep{upsilon}. \textit{Upsilon} is a general purpose random forest classifier trained using OGLE and EROS-2 data \citep{upsilon}. For each ASAS-SN light curve, \textit{Upsilon} calculates a set of features and provides a general variability classification. \textit{Upsilon} also classifies certain variability classes into sub-classes (as an example, RR Lyrae are further classified as RRab, RRc, RRd and RRe). We consider sub-classes for the RR Lyrae, Cepheid and eclipsing binary classifications, but merge the other variability types with \textit{Upsilon} sub-classes into a single class.

After classifying the ASAS-SN variable sources with \textit{Upsilon}, we visually verified each classification. The RR Lyrae, Cepheids and DSCT candidates were examined by at least 2 authors, while the remainder of the variability types were examined by only one author. Variables that pass the visual review at this stage are published in the catalog and the remainder were saved for the next stage of the classification process. Through this process of visual review, we identified 22,809 periodic variables. To characterize these variables, we defined the amplitude $A$ as the difference in the 2.5$^{th}$ and 97.5$^{th}$ percentiles in magnitude to mitigate the effects of outliers.

Variables were also tagged as `variable/not irregular', `variable/irregular' or `bad' during the inspection process. Those tagged as `variable/not irregular' were saved for the next stage of classification. Variables that were tagged as `bad' generally had issues with photometry (bleeding trails, saturation, etc.) or astrometry (light curve extracted from the wrong position). These variables were removed from our working list. A further $\sim 37,000$ candidates were tagged as `variable/irregular' through this visual review.

Based on these reviews, we found that \textit{Upsilon} achieves sufficient precision for well-defined variability types (e.g, RR Lyrae, Mira, eclipsing binaries and $\delta$ Scuti) but fares poorly for other types (Table \ref{tab:upsilonprec}). The precision values in Table \ref{tab:upsilonprec} were calculated as the ratio of the size of the positively reviewed sample to the size of the sample classified by \textit{Upsilon} for any given variability type. Most of the false positives at this stage can be attributed to irregular variables with aliased periods and noisy data.

We aim to classify these variables according to the variability types defined by the VSX \citep{2006SASS...25...47W} and GCVS \citep{2017ARep...61...80S} catalogs. Most of the variability types defined in the \textit{Upsilon} classifier do not strictly map onto a VSX/GCVS type, therefore we changed the \textit{Upsilon} classifications based on the VSX/GCVS definitions. \textit{Upsilon} classifies eclipsing binaries into three types: EC (contact binaries), ESD (semi-contact binaries) and ED (detached binaries). EC classifications are changed to the `EW' VSX type (W UMa type binaries), while the majority of the ESD classifications are assigned to the `EA$|$EB' (Algol-type or $\beta$-Lyrae type) VSX class. Most variables classified as ESD with periods $P<0.4$ d are EW types, and were reclassified accordingly. ED classifications were changed to the `EA' VSX type (Algol-type binaries).

Cepheids are classified as fundamental mode, first overtone and type II Cepheids by \textit{Upsilon}. We visually reviewed each classification to sort them into the VSX types DCEP (fundamental mode), DCEPS (first overtone), CWA/CWB (Type II: BL Herculis/W Virginis types), AHB1 (`Above Horizontal Branch' variables of subtype 1) and ROT (rotational variables). Some $\sim14\%$ of the variables classified as Cepheids by \textit{Upsilon} are spotted variables showing evidence of rotational modulation. CWA and CWB variables are separated by period, with CWA (CWB) variables having $P>8$ d ($P<8$ d). AHB1 variables have highly asymmetric light curves with large amplitudes that resemble the light curves of RRab variables, allowing them to be separated from short period fundamental mode Cepheids (DCEP).

\textit{Upsilon} assigns $\delta$-Scuti variables to the DSCT class. We further categorize this class by selecting high-amplitude $\delta$-Scuti (HADS) variables based on the amplitude cut $A>0.15$ mag. Semi-regular variables are assigned the VSX types of SR and SRS based on their period. SRS variables are short period semi-regular variables with $P<30$ d, while semi-regular SR variables have $P>30$ d. 

Mira variables (VSX type `M') in our sample frequently dip below our detection limits ($V\gtrsim17$), hence their ASAS-SN amplitudes can be less than the minimum V-band amplitude of $A=2.5$ mag used by the VSX/GCVS catalogs to distinguish between high amplitude semi-regular variables and Miras. This can also lead to incorrect periods in a few cases where the light curve is inadequately sampled. We have visually reviewed the Miras to verify the shapes of their light curves and periods but we caution that the true period might differ from what is reported in this catalog.

To classify the $\sim 37,000$ irregular variables, we utilize the 2MASS and APASS magnitudes following the criteria given in Table \ref{tab:irrclass}. The VSX classes `L' and `L:' are assigned to the irregulars with red $J-K_s$/$B-V$ colors and no infrared excess, while the VSX class `YSO' (young stellar object) is assigned to the irregulars with faint infrared magnitudes ($J>10$), intermediate to red colors and an infrared excess. Blue irregulars with $J-H <0.2$ and $B-V<0.2$ are classified as Be stars in outburst (VSX class `GCAS'). Candidates that met this criteria were visually reviewed to confirm the presence of outbursts. We were able to identify 79 GCAS candidates. The irregulars that do not fit into these classes are assigned the generic `VAR' (variable) class. These irregular variables were also added to the catalog at this stage.
\begin{table}
	\centering
	\caption{Performance of the \textit{Upsilon} classifier}
	\label{tab:upsilonprec}
	\begin{tabular}{lc} 
		\hline
		Class & Precision\\
		\hline
        $\delta$ Scuti & 68$\%$ \\
		RR Lyrae & 83$\%$ \\
		Cepheids & 10$\%$\\
		Eclipsing Binaries & 63$\%$ \\
		Semi-regular variables & 36$\%$ \\  
		Mira & 75$\%$ \\        
		\hline
	\end{tabular}

\end{table}   

\begin{table}
	\centering
	\caption{Sub-classes assigned to the irregular variables}
	\label{tab:irrclass}
    \begin{threeparttable}
      \begin{tabular}{lc} 
          \hline
          Class & Criteria\\
          \hline
          L & $J<10$ \& $J-K_s >1.1$ \& $B-V>1.5$ \\
          L: & $J<10$ \& $J-K_s >1.1$ \& $B-V$ not available \\
          YSO & $J>10$ \& $J-K_s <1.1$ \\  
          \space & $J>10$ \& $J-K_s >1.1$ \& $B-V<1.5$ \\ 
          GCAS &  $J-H <0.2$ \& $B-V<0.2$ + Visual Review \\
          VAR & Remainder of the sample \\
          \hline        
      \end{tabular}
    \begin{tablenotes}
      \small
      \item $J$, $H$ and $K_s$ are the 2MASS magnitudes, while the $B$ and $V$ magnitudes are from APASS.
    \end{tablenotes}
  \end{threeparttable}      
\end{table}   

\subsection{Random Forest Classification}
\label{rfc}
Next we built our own classifier based on the ASAS-SN data using the \verb"scikit-learn"  \citep{2012arXiv1201.0490P} implementation of the random forest classifier. We set the number of decision trees in the forest to be \verb"n_estimators=700" and the number of features considered at each branching point to be \verb"max_features=`log2'". The base 2 logarithm of the total features passed onto the random forest classifier is widely considered to be the optimal choice for this parameter. To prevent the forest from over-fitting the data, we pruned the trees at a maximum depth of \verb"max_depth=12", set the number of samples needed to split a node to \verb"min_samples_split=10" and set the number of samples at a leaf node to be \verb"min_samples_leaf=10". Our training sample consists of variables that passed the \textit{Upsilon} classification and visual review process. There is a distinct imbalance of variables amongst the different classes. The irregular variables ($\sim10^4$ sources) outnumber all the other variability types. The smallest classes, like the DCEP variables, consist of a meager $\sim10^2$ sources.  An imbalanced training set can cause over-fitting, hence we assign weights to each class in the training set by initializing \verb"class_weight=`balanced_subsample'". These parameters were optimized using cross-validation.

\begin{table*}
	\centering
	\caption{Variability features used to train the random forest classifier}
	\label{tab:features}
	\begin{tabular}{lcc} 
		\hline
		Feature & Description & Reference \\
		\hline
        logP & Base 10 logarithm of the period & -  \\
        $J-H$ & 2MASS $J-H$ color  & \citet{2006AJ....131.1163S} \\
        $A$ & Amplitude of the light curve & - \\ 
        $M_{s}$ & Skewness of the magnitude distribution & -\\ 
        $M_{k}$ & Kurtosis of the magnitude distribution & -\\          
        $T_m$ & M-test statistic & \citet{2006AJ....132.1202K} \\   
        IQR & Difference between the 75$^{th}$ and 25$^{th}$ percentiles in magnitude & -\\ 
        $a_{42}$ & Ratio between the Fourier components $a_{4}$ and $a_{2}$ & -\\ 
        $b_{42}$ & Ratio between the Fourier components $b_{4}$ and $b_{2}$ & - \\   
        $R_{31}$ & Ratio between the amplitudes of the $3^{rd}$ and $1^{st}$ harmonics & - \\  
        $R_{21}$ & Ratio between the amplitudes of the $2^{nd}$ and $1^{st}$ harmonics & - \\   
        $\Phi_{42}$ & Difference between the phase angle of the $4^{th}$ and $2^{nd}$ harmonics & -\\ 
        $\Phi_{31}$ & Difference between the phase angle of the $3^{rd}$ and $1^{st}$ harmonics & -\\ 
        $A_{HL}$ & Ratio of magnitudes brighter or fainter than the average & \citet{upsilon} \\
        rms & Root-mean-square statistic for the light curve  & -  \\ 
        K & Stetson K variability index  & \citet{1996PASP..108..851S}  \\ 
		\hline
	\end{tabular}
\end{table*} 

In addition to tuning the random forest, one must choose the features of the data that may distinguish between the variable classes. We use a set of 16 features that have commonly been used in similar variability classification problems (see \citealt{2018MNRAS.475.2326P}, and references therein). For the periodic variables, we fit a Fourier model,
\begin{equation}
    M(\phi)=m_0 + \sum_{i=1}^{4} \left(a_isin(2\pi i\phi)+b_icos(2\pi i\phi)\right)
	\label{eq:ffit}
\end{equation} where $m_0$ is fixed as the median magnitude for each source and $\phi$ is the phase at a given epoch. The amplitude of the $i^{th}$ harmonic is given as $A_i={(a^2_i+ b^2_i)^{{1}/{2}}}$ and the phase angle is $\Phi_i =\tan^{-1}\space({b_i}/{a_i})$. As variability features, we use the amplitude ratios of the harmonics defined as $R_{ij}=A_i/A_j$, the ratios of the Fourier components defined by $a_{ij}=a_i/a_j$ and $b_{ij}=b_i/b_j$, and the difference in phase angle between two harmonics defined by $\Phi_{ij} =\Phi_{i}-\Phi_{j}$. The complete list of features is summarized in Table \ref{tab:features}.

\subsubsection{Training and Evaluating the Classifier}

We split the classified variables (\textit{Upsilon} + visual review) into separate lists for training ($80\%$) and testing ($20\%$). To minimize class confusion, we merged variability types having many sub-classes into a single class. The final list of classes is: DSCT, CEPH, RRab, RRc, M, SR, IRR, EA, EA|EB and EW. Variability types with $<50$ members (i.e ROT, RRD, etc.) were dropped from both the training and test lists. We hope to revisit such classes at a later stage of the project where we supplement our training set with light curves for all the known, bright variables.

\begin{figure}
	\includegraphics[width=0.5\textwidth]{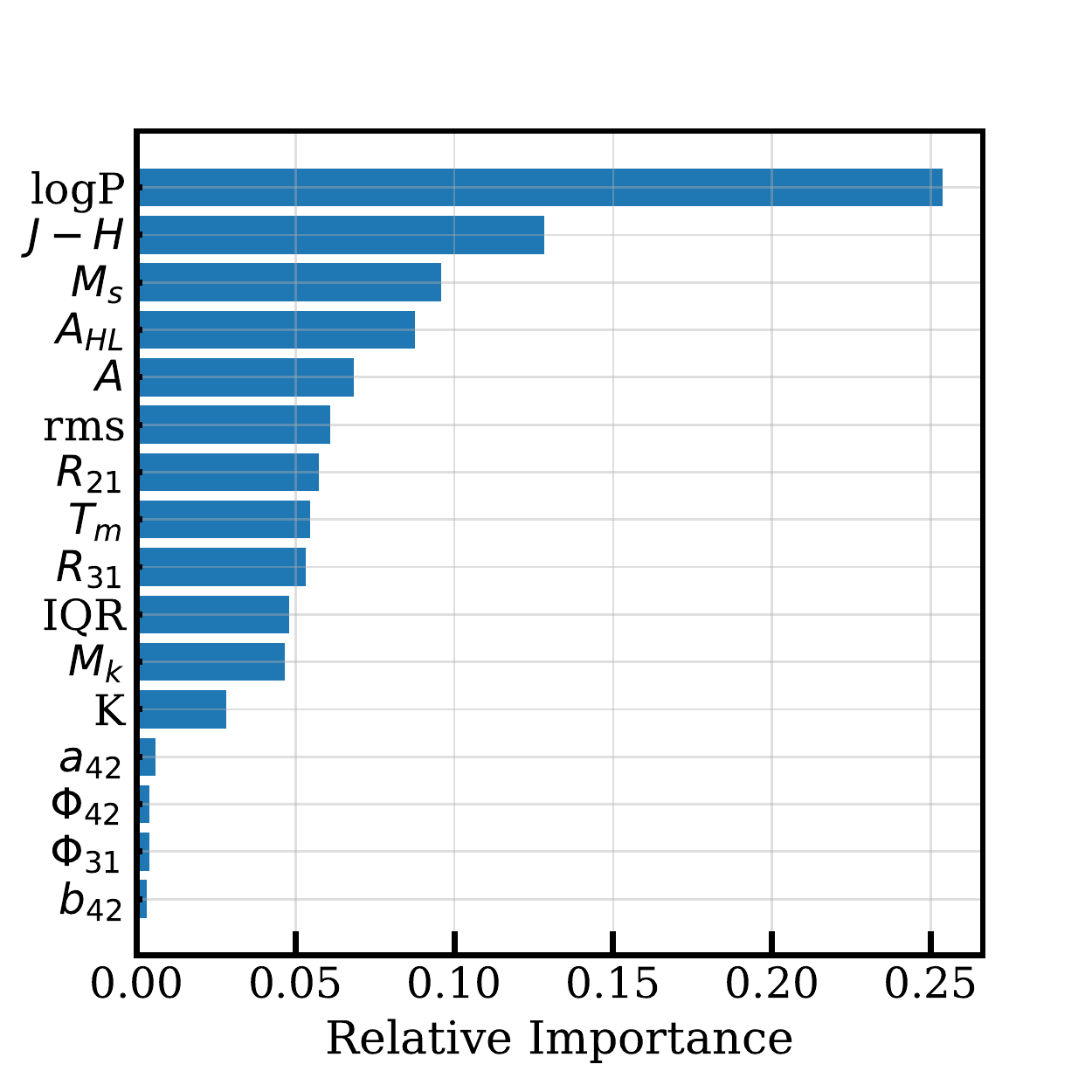}
    \caption{A bar chart illustrating the relative importance of the features listed in Table \ref{tab:features} that were used to train the Random Forest Classifier.}
    \label{fig:fig3}
\end{figure}

\begin{figure*}
	\includegraphics[width=\textwidth]{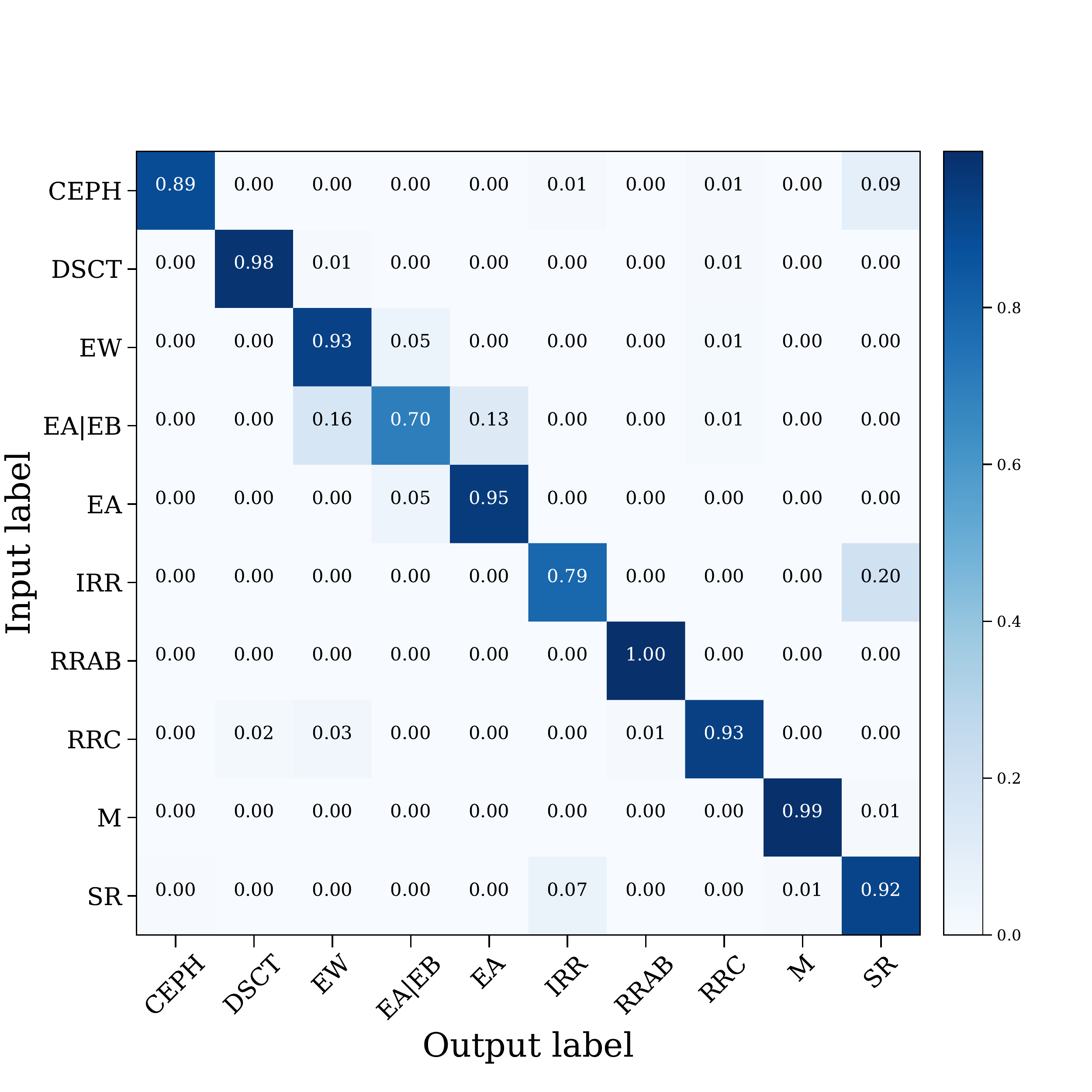}
    \caption{The normalized confusion matrix derived from the trained random forest classifier for the ASAS-SN variables. The y-axis corresponds to the `input' classification given to a variable, while the x-axis represents the `output' prediction obtained from the trained random forest model.}
    \label{fig:fig4}
\end{figure*}

Figure \ref{fig:fig3} highlights the feature importances derived from the trained random forest model. The period and color provides the greatest discrimination between the variable sources, with a combined importance of 38.8$\%$. Several features ($a_{42}$, $b_{42}$, $\Phi_{42}$, $\Phi_{31}$ and K) do not discriminate significantly between the variables (importances $\lesssim 5\%$). Figure \ref{fig:fig4} illustrates the random forest model's ability to classify new objects. A perfect model would correctly classify all the objects, leading to a perfectly diagonal confusion matrix. Class confusion is greatest between the sub-classes of the eclipsing binaries, but this confusion does not significantly extend to mixing the eclipsing binaries with the non-binary classes. Semi-regulars and irregulars are also frequently confused, which is to be expected given the somewhat hazy distinction between them.

We can evaluate the overall results based on the precision, \begin{equation}
    P=\frac{T_p}{T_p + F_p} \,,
	\label{eq:prec}
\end{equation} and recall, \begin{equation}
    R=\frac{T_p}{T_p + F_n}\,,
	\label{eq:rec}
\end{equation} where $T_p$ and $F_p$ ($T_n$ and $F_n$) are the numbers of true and false positives (negatives). The precision and recall values for each class used in the random forest classifier are shown in Table \ref{tab:scores}. The overall precision and recall parameters of 89$\%$ and 85$\%$ respectively were calculated using a weighted average.

The final random forest model was trained using the entire dataset with the same initialization conditions as those used to test and verify the model. We used the trained random forest model to re-classify $\sim 6,000$ variable sources that were potentially misclassified by the \textit{Upsilon} classifier and an additional $\sim 4,000$ new variable candidates identified in the interim. All the classifications were reviewed by at least a single author to ensure their accuracy and to verify the quality of the light curves before including these sources in the catalog.

\begin{table}
	\centering
	\caption{Precision and Recall Values by class.}
	\label{tab:scores}
	\begin{tabular}{lcc} 
		\hline
		Class & Precision & Recall\\
		\hline
		EA & 63$\%$ & 97$\%$ \\
		EA|EB & 55$\%$ & 70$\%$ \\
		EW & 93$\%$ & 93$\%$ \\
        $\delta$ Scuti & 91$\%$ & 99$\%$ \\
		RRab & 98$\%$ & 100$\%$ \\
		RRc & 90$\%$ & 92$\%$\\
		Cepheids & 69$\%$ & 89$\%$ \\
		Mira & 89$\%$ & 99$\%$\\     
		Semi-regular variables & 63$\%$ & 99$\%$ \\          
		Irregular variables & 97$\%$ & 79$\%$\\          
		\hline
	\end{tabular}
\end{table}

\section{The New Variable Stars}
\label{catalog}
The complete catalog of 66,179 new variables is available at the ASAS-SN Variable Stars Database (\href{https://asas-sn.osu.edu/variables}{https://asas-sn.osu.edu/variables}) along with the V-band light curves for each source. Table \ref{tab:var} lists the number of sources of each variability type in the catalog. This catalog can be downloaded in its entirety at the ASAS-SN Variable Stars Database and is also included in the VSX catalog of variables. This database will continue to be updated as we discover new variables and reclassify existing ones. We intend to populate this database with the light curves of all the known variables in our data over the next year. Our goal with the ASAS-SN Variable Stars Database is to provide a homogeneous set of light curves for all bright, known variable sources (V$\lesssim17$ mag).

\begin{table*}
	\centering
	\caption{New ASAS-SN variables by type}
	\label{tab:var}
	\begin{tabular}{lcr} 
		\hline
		Class & Description & Number Discovered\\
		\hline
		EA & Detached Algol-type binaries & 1530 \\
		EA|EB & Algol-type or $\beta$ Lyrae-type binaries & 529\\
		EW & W Ursae Majoris type binaries & 2427 \\
        DSCT & $\delta$ Scuti type variables & 22 \\
        HADS & High amplitude $\delta$ Scuti type variables& 548 \\           
		RRAB & RR Lyrae variables with asymmetric light curves & 4820 \\
		RRC & First Overtone RR Lyrae variables & 653\\
		RRAB/BL & RRAB variables showing the Blazhko effect & 40 \\
		RRC/BL & RRC variables showing the Blazhko effect & 7\\        
        RRD & Double Mode RR Lyrae variables & 39 \\
        AHB1 & Above Horizontal Branch Subtype 1 (XX Vir) Variables & 24 \\         
		DCEP & $\delta$ Cephei-type/classical Cepheid variables & 160\\
		DCEPS & First overtone Cepheid variables  & 84 \\
		CWA & W Virginis type variables with $P>8$ d & 22 \\ 
		CWB & W Virginis type variables with $P<8$ d (BL Herculis variables) & 44 \\   
		RVA & RV Tauri variables (Subtype A) & 3 \\         
		M & Mira variables & 1853\\  
		SRS & Short period semi-regular variables with $P<30$ d  & 162\\           
		SR & Long period semi-regular variables with $P>30$ d & 13844 \\ 
		L & Red irregular variables & 18768\\ 
		L: & Irregular variables with missing B-V colors & 10620\\  
		YSO & Young Stellar Objects & 4215\\ 
		GCAS & Rapidly rotating early type stars & 93\\         
		ROT & Spotted Variables with rotational modulation & 662\\         
		VAR & Variable star of unspecified type & 5010\\            
		\hline
	\end{tabular}
\end{table*}

We note that the ASAS-SN catalog is not complete - this was not an endeavor to do a statistically well-defined, all-sky search for variable stars, but rather a classification of variable sources that were flagged during our search for supernovae. Nevertheless, this catalog is likely to contain a significant fraction of all previously undiscovered, high amplitude variables at these magnitudes. High amplitude variables are more likely to be flagged as variables during the search for supernovae since these sources stand out from other low amplitude variables in the difference images. The estimated variability amplitudes can be underestimated due to crowding and blending, particularly in the Galactic plane.

\begin{figure*}
	\includegraphics[width=1.1\textwidth]{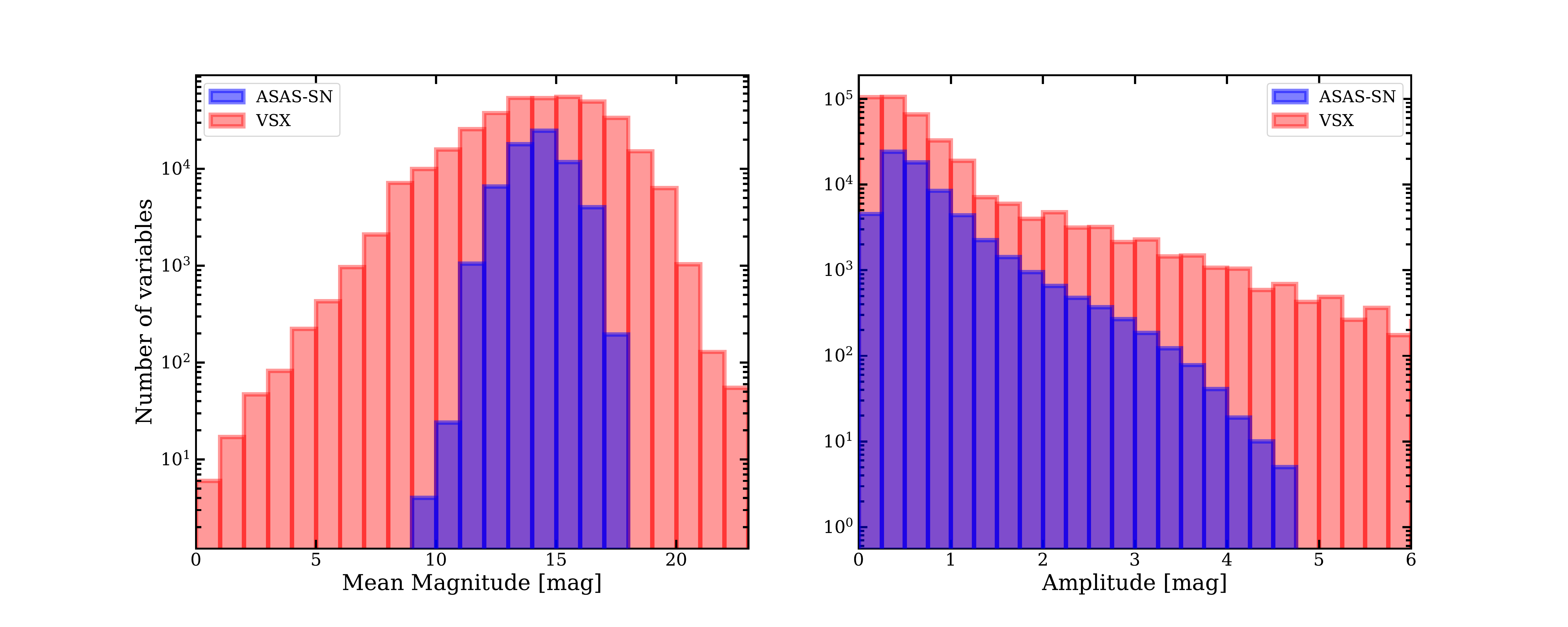}
    \caption{Histograms of the mean V-band magnitudes (\textit{left}) and amplitudes (\textit{right}) of the $\sim 66,000$ new ASAS-SN variables (blue) as compared to the $\sim 370,000$ known variables from the VSX catalog (red) with amplitudes $A>0.05$ mag. The average variable has a mean magnitude of ${V}\sim 14.3$ mag and mean amplitude of $A\sim 0.69$ mag.}
    \label{fig:fig5}
\end{figure*}

\begin{figure*}
	\includegraphics[width=1.1\textwidth]{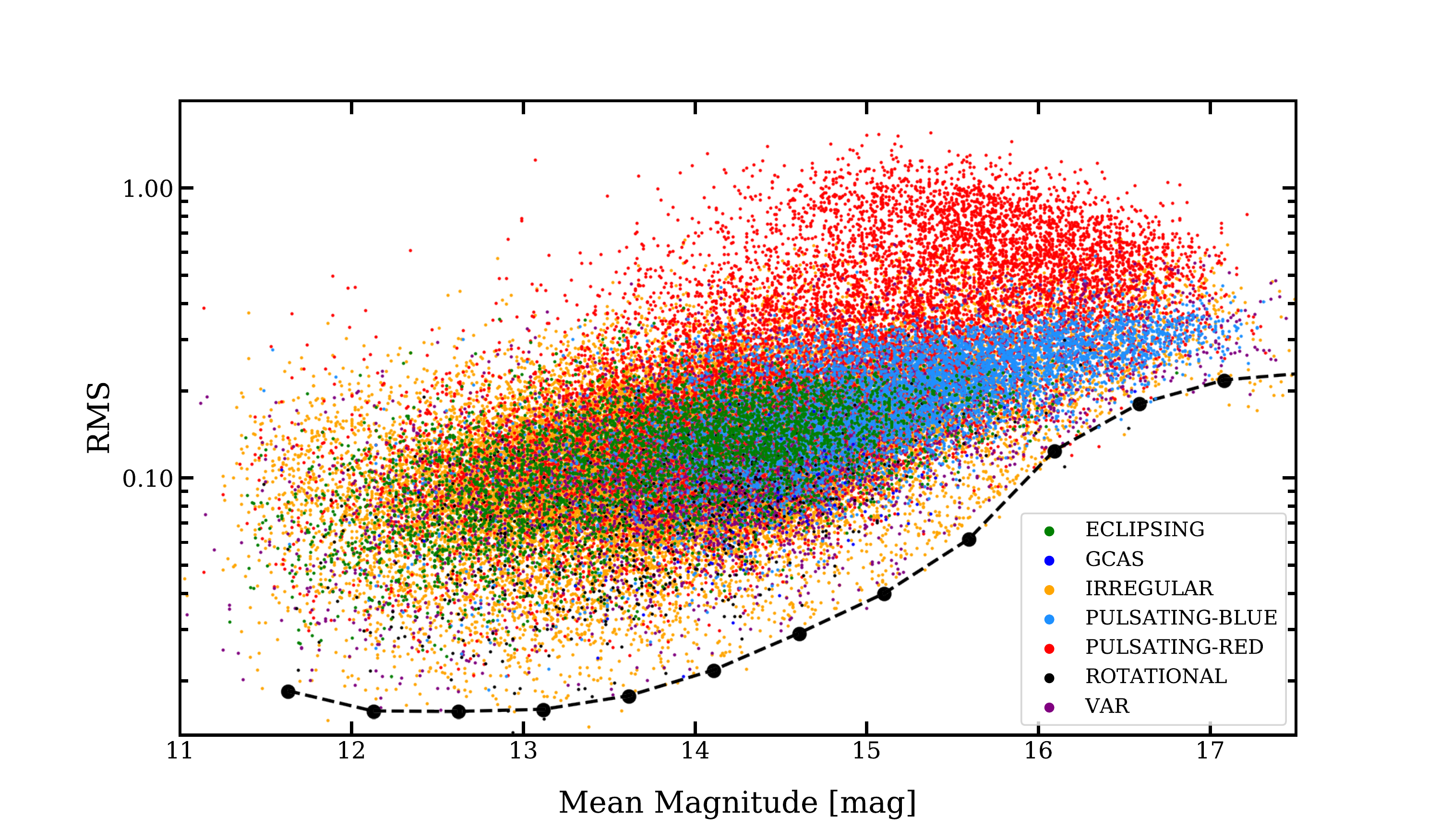}
    \caption{Root-mean-square variability as a function of mean magnitude for the new variable sources colored by the variability groups described in the legend. The median RMS of the APOGEE light curves used in $\S2.1$ is shown in black.}
    \label{fig:fig6}
\end{figure*}

Figure \ref{fig:fig5} shows the distribution of these new variables in mean brightness and amplitude (blue) as compared to the $\sim 370,000$ other pulsating and eclipsing variables in the VSX catalog with amplitudes $A>0.05$ mag (red). The average variable in our catalog has a mean brightness of $\overbar{V}\sim 14.3$ mag and a mean amplitude of $A\sim 0.69$ mag. Figure \ref{fig:fig6} shows the root-mean-square (RMS) of the light curves as a function of their mean brightness. Not surprisingly, fainter variables tend to have higher amplitudes. The high amplitude variables at bright magnitudes will have been found by shallower searches like ASAS. Our new catalog increases the number of known variables with magnitude $13<V<16$ by$\sim 30\%$, largely by filling in the lower amplitude variables missed by (presumably) shallower surveys.

To examine the distribution of these variables in 2MASS color-color and color-period space, we grouped the variables into 7 categories. Blue pulsators are the RR Lyrae, Cepheid (including type II Cepheids), $\delta$ Scuti and AHB1 variables. Red pulsators are the Mira and semi-regular variables. The irregular variables are the L, L: and YSO sub-types. Eclipsing variables contain the EW, EA|EB and EA classes. The GCAS, Rotational (ROT) and VAR categories directly map onto their respective VSX classes. 

\begin{figure*}
	\includegraphics[height=0.90\textheight]{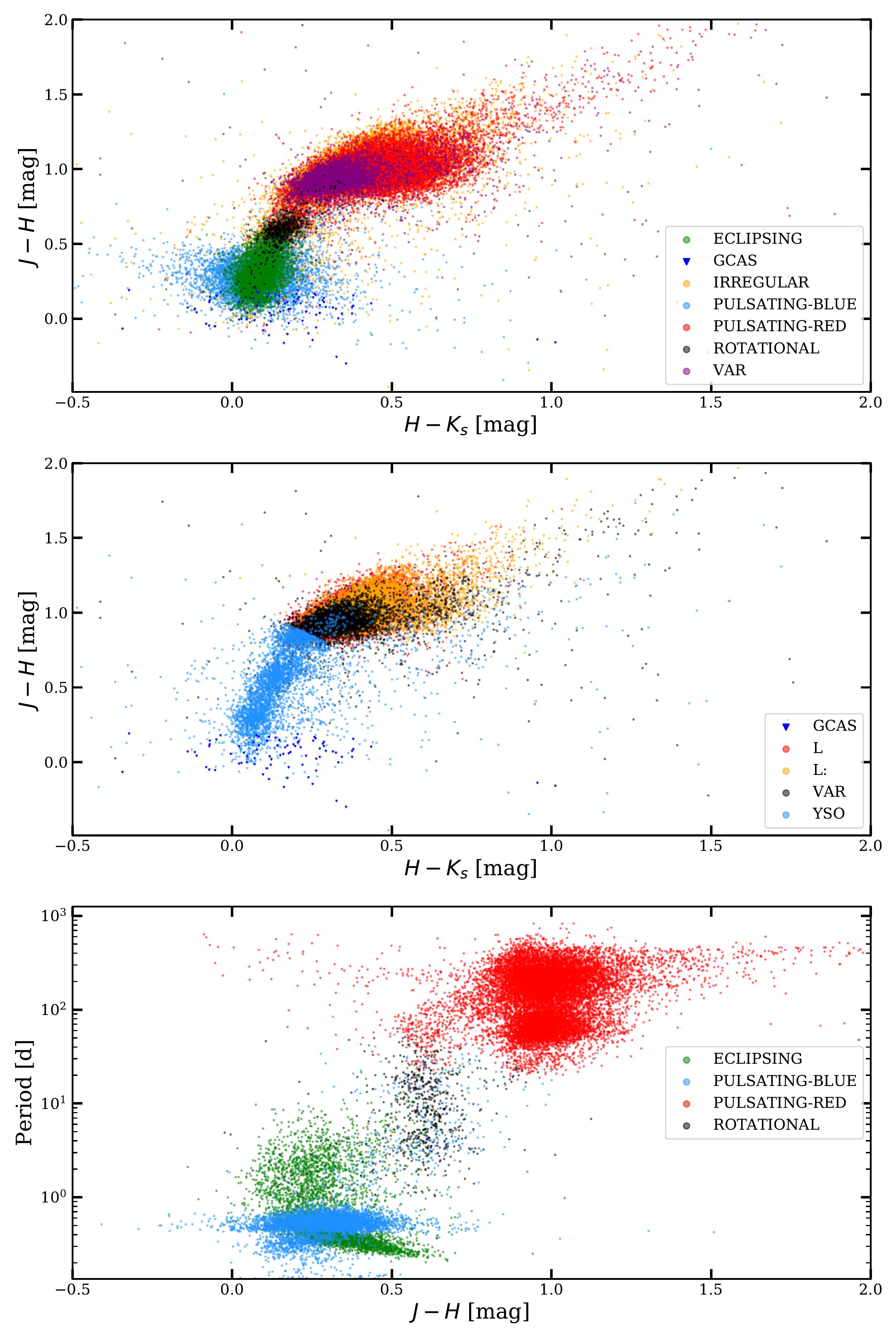}
    \caption{\textit{Top:} The 2MASS color-color diagram for all the variables in our sample colored by the variability groups described in the legend. \textit{Middle:} The 2MASS color-color diagram for the irregular variables in our sample colored by the variability groups described in the legend. \textit{Bottom:} Period-color diagram for the periodic variables in our sample.}
    \label{fig:fig7}
\end{figure*}

Figure \ref{fig:fig7} shows the distribution in 2MASS color ($J-H$, $H-K_s$) and period ($J-H$, P). There is a clear dichotomy between the red and blue pulsators in our sample. Most of the eclipsing systems are composed of blue, likely main sequence stars. The generic VAR category mostly overlaps with the red pulsators, suggesting that a large fraction of these are actually red irregular variables. GCAS variables mostly occupy the lower left quadrant, while rotational variables are scattered amongst the blue and red pulsators. The distinction between the different sub-types assigned to the ASAS-SN sample of irregulars in the 2MASS color-color space is shown in the middle panel. 

Red pulsators have the largest periods and are clustered at the top right quadrant of the bottom panel. Sub-structure is evident in the distribution of the blue pulsators, highlighting the different classes of variables that comprise this group. The eclipsing binaries overlap the distribution of blue pulsators having similarly short periods. Eclipsing binaries span a large range of periods from relatively short periods ($\sim 0.2$ d) to large periods ($\sim 10^4$ d). Long period eclipsing binaries with periods $>10^2$ d are rare in our sample.

\begin{figure*}
	\includegraphics[width=0.99\textwidth]{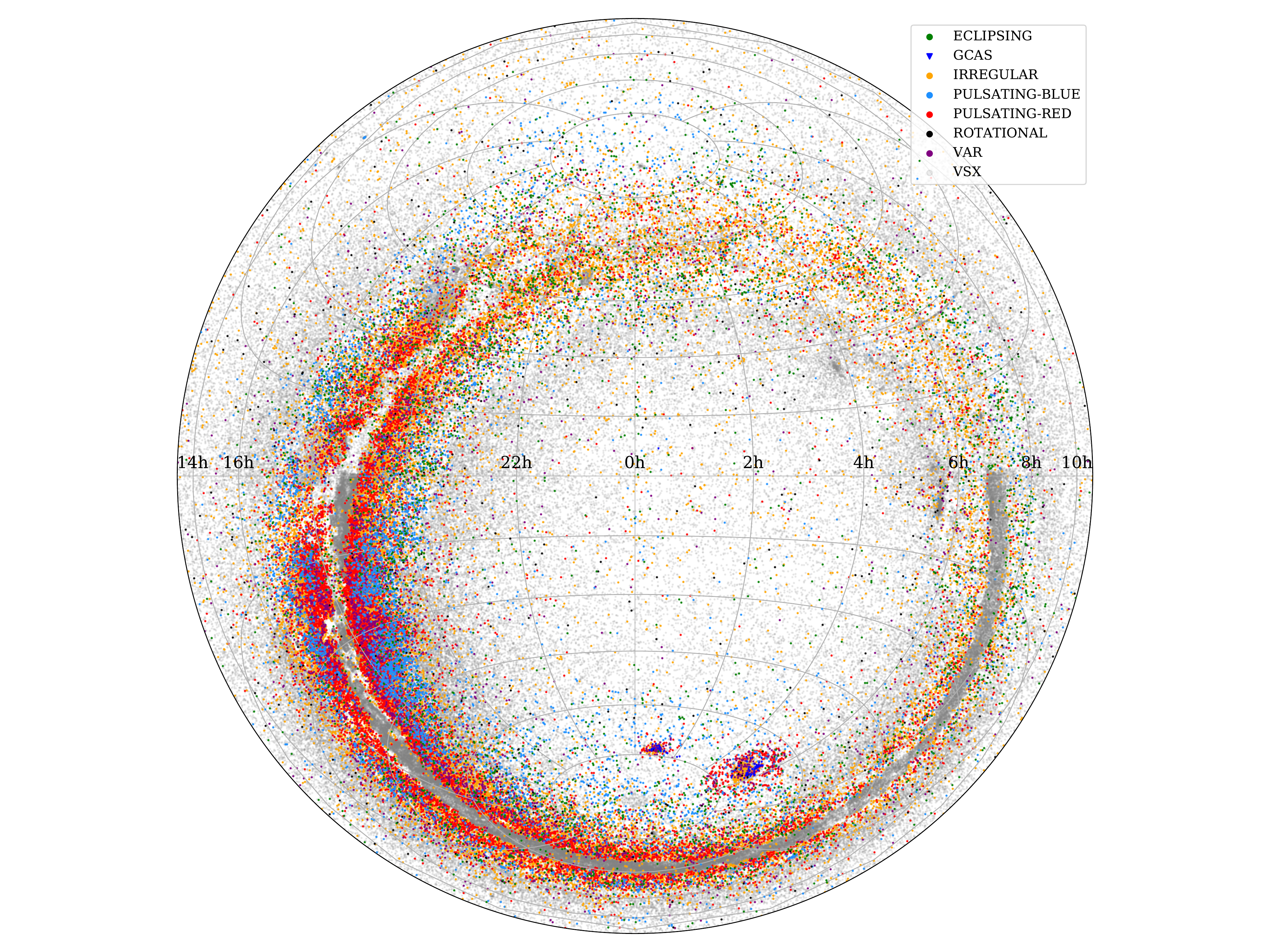}
    \caption{Spatial distribution of the $\sim 66,000$ new ASAS-SN variables in Equatorial coordinates. Grey points denote the $\sim 370,000$ known variables from the VSX catalog with V$<17$ mag and $A>0.05$ mag.}
    \label{fig:fig8}
\end{figure*}

Figure \ref{fig:fig8} shows the spatial distribution of the ASAS-SN variables in Equatorial coordinates. Most of our new variables are just above or below the Galactic disk. Many surveys avoided lower Galactic latitudes (e.g., \citealt{2014ApJS..213....9D}) and the few surveys of the Galactic plane tended to cover only a few degrees in latitude (e.g., \citealt{2012AN....333..706H,2014Msngr.155...29H}).

\subsection{Pulsating Variables}
Our catalog is largely ($\sim 95 \%$) composed of pulsating variables, and from the $\sim 61,000$ pulsating variables discovered by ASAS-SN, $\sim 37 \%$ are periodic variables. Here we discuss the classes of pulsating variables in the ASAS-SN catalog of variable stars and provide examples of the light curves for these classes of variables.

\subsubsection{$\delta$ Scuti}
\begin{figure*}
	\includegraphics[width=0.99\textwidth]{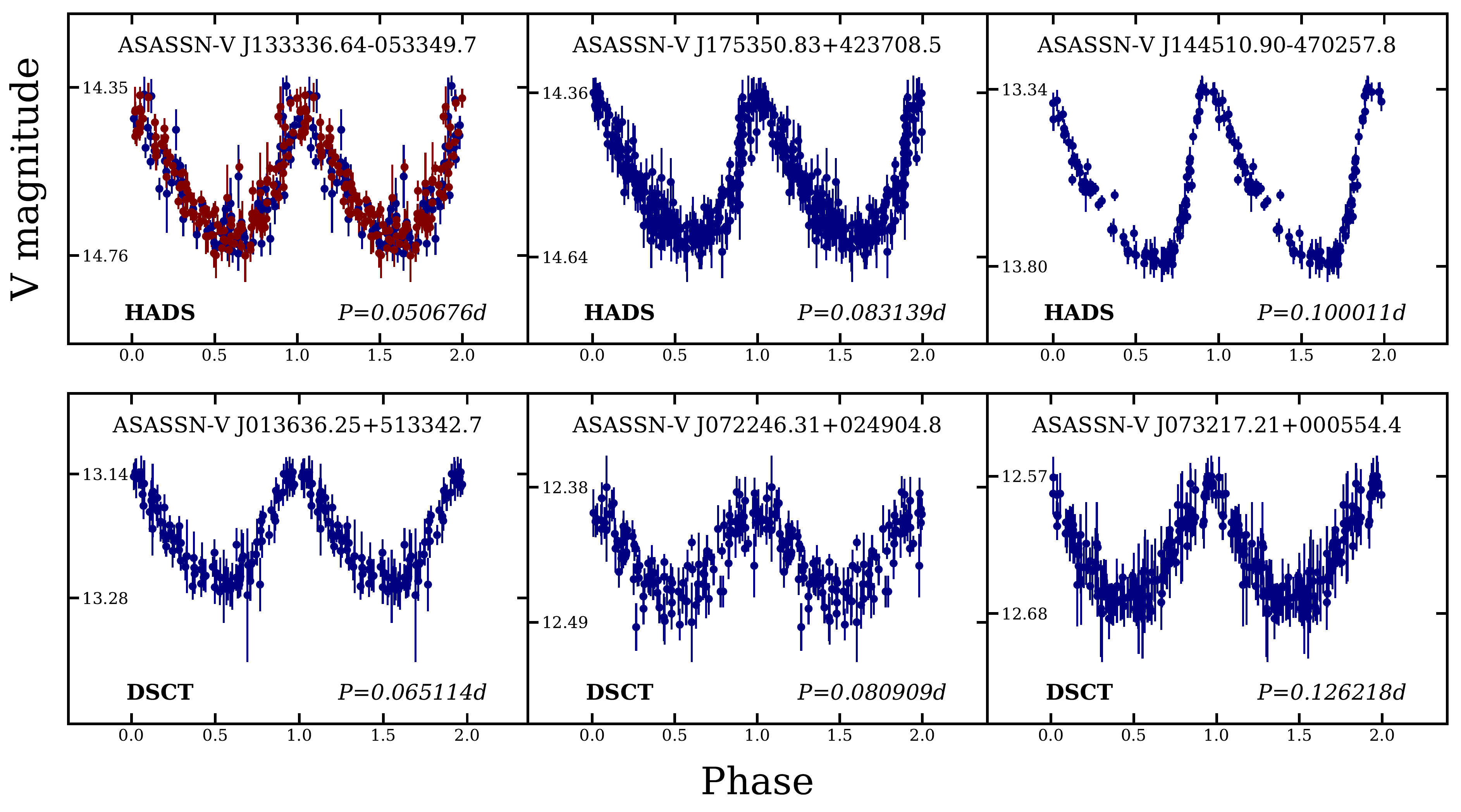}
    \caption{Phased light curves for examples of the DSCT and HADS VSX classes. The light curves are scaled by their minimum and maximum V-band magnitudes with different variable classes in each row. Different colored points correspond to data from the different ASAS-SN cameras.}
    \label{fig:fig9}
\end{figure*}

$\delta$ Scuti stars are high frequency pulsators (P$<0.3$ d) located at the lower end of the instability strip \citep{1979PASP...91....5B}. The $\delta$ Scuti variables are commonly sub-divided based on their amplitude. High amplitude $\delta$ Scuti variables (HADS) have amplitudes ($A>0.15$ mag) and are easily discovered by sky surveys similar to ASAS-SN. $\delta$ Scuti variables (DSCT) have amplitudes ($A<0.15$ mag) and are somewhat less commonly found due to the selection effect caused by their low amplitudes. This statement is reflected in the distribution of HADS/DSCT variables in our catalog: $\sim 96\%$ are HADS variables. Examples of the DSCT and HADS classes are shown in Figure \ref{fig:fig9}.

\subsubsection{RR Lyrae and AHB1/XX Vir Variables}

\citet{2018RNAAS...2...18J} announced the discovery of 4,880 RR Lyrae from the first stage of our variability classification process. Since then, we have discovered an additional 679 RR Lyrae in our data. Our catalog of RR Lyrae stars largely consists of the RRAB ($\sim 87\%$) and RRC ($\sim 12\%$) types. A small fraction consists of double-mode RRD stars and RRAB/RRC stars showing the Blazhko effect \citep{1907AN....175..325B}. 

A common issue that arises in the classification of RRC variables is cross-contamination with short period contact binaries (EW). This is especially problematic for sources with amplitudes $0.3<A<0.5$ mag and periods $0.44<P<0.82$ d \citep{2014ApJS..213....9D}. In this parameter space, RRC variables can have almost symmetric light curves with morphologies similar to those of EW binaries phased with half their true period. EW binaries tend to have the light curve skewness parameter $M_s>0$, while most RRC variables have $M_s<0$. One can potentially use this knowledge to reduce cross-contamination between these two variability types, but ultimately, one has to resort to spectroscopic follow-up to distinguish between these two variability types in the small number of cases where the light curves are photometrically ambiguous.

During the process of visual review, we identified 39 double-mode RR Lyrae candidates. \citet{2014PASP..126..509P} studied double-mode radial pulsations amongst RR Lyrae stars and produced a Petersen diagram for RRD stars with the majority of their sample having $0.742<{P_{1O}}/{P_{FO}}<0.748$. Our RRD candidates were searched for secondary periods, and those with period ratios $0.73<{P_{1O}}/{P_{FO}}<0.77$ were classified as RRD stars. We chose this range in ${P_{1O}}/{P_{FO}}$ to account for variations observed in RRD populations \citep{2009A&A...494L..17O} and parameter dispersions due to noise. The periods reported for the RRD stars are the best period derived for each light curve, where in most cases this is the period of the fundamental mode. We also identified 40 (7) RRAB (RRC) stars showing the Blazhko effect. These sources were classified based on the scatter around the primary maximum, which is indicative of the Blazhko effect.

AHB1 variables are evolved horizontal branch pulsators with periods $0.8<P<3$ d and asymmetric light curves resembling those of RRAB stars \citep{2006ARA&A..44...93S}. These variables are distinguished from low period fundamental mode Cepheids by their low rise times to maximum and high visual amplitudes. We identify 24 candidate AHB1 variables.

\begin{figure*}
	\includegraphics[height=0.90\textheight]{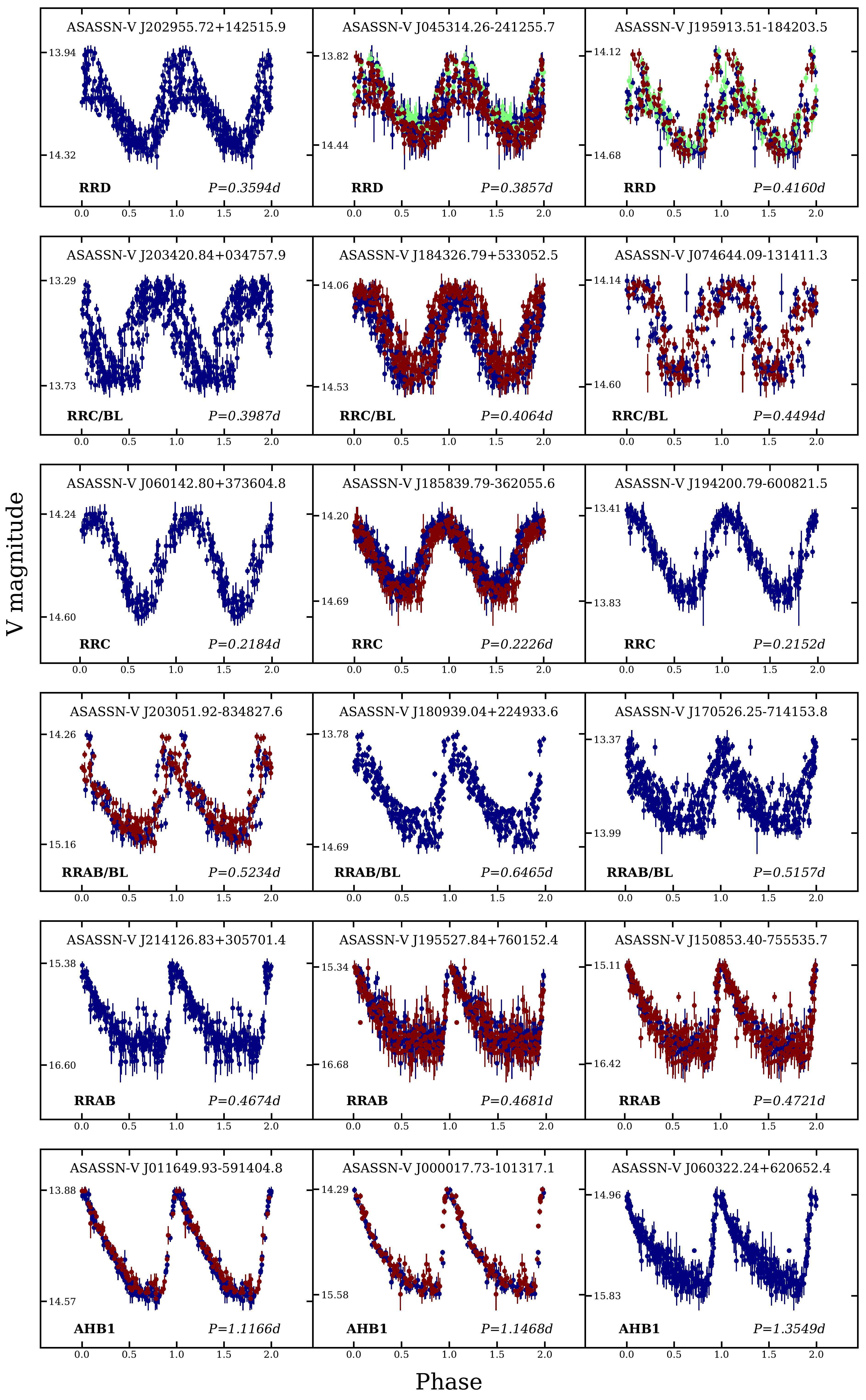}
    \caption{Phased light curves for examples of the RR Lyrae and AHB1 classes. The format is the same as for Fig. \ref{fig:fig9}.}
    \label{fig:fig10}
\end{figure*}

Examples of the RR Lyrae and AHB1 classes are shown in Figure \ref{fig:fig10}.

\subsubsection{Cepheids}
Large numbers of these variables have been discovered over the years since the first Cepheids were discovered in the Magellanic clouds by \citet{1908AnHar..60...87L}. For example, the current OGLE catalog of Cepheids in the Magellanic Clouds boasts close to 10,000 Cepheids \citep{2017AcA....67..103S}. In the Milky Way, only $\sim1000$ Cepheids are currently known \citep{2011A&A...530A..76W}. Fundamental mode Cepheids (DCEP) are the archetypal classical Cepheids obeying a period-luminosity relation \citep{1912HarCi.173....1L}. DCEP variables have unique light curve morphologies as a function of their period. First overtone Cepheids (DCEPS) have almost symmetrical light curves with periods $P<7$ d and amplitudes $A<0.5$ mag. We identified 160 fundamental mode and 84 first overtone Cepheids in our data. Over $\sim 80\%$ of the ASAS-SN DCEP and DCEPS variables are towards the Galactic disk.

\begin{figure*}
	\includegraphics[width=0.95\textwidth]{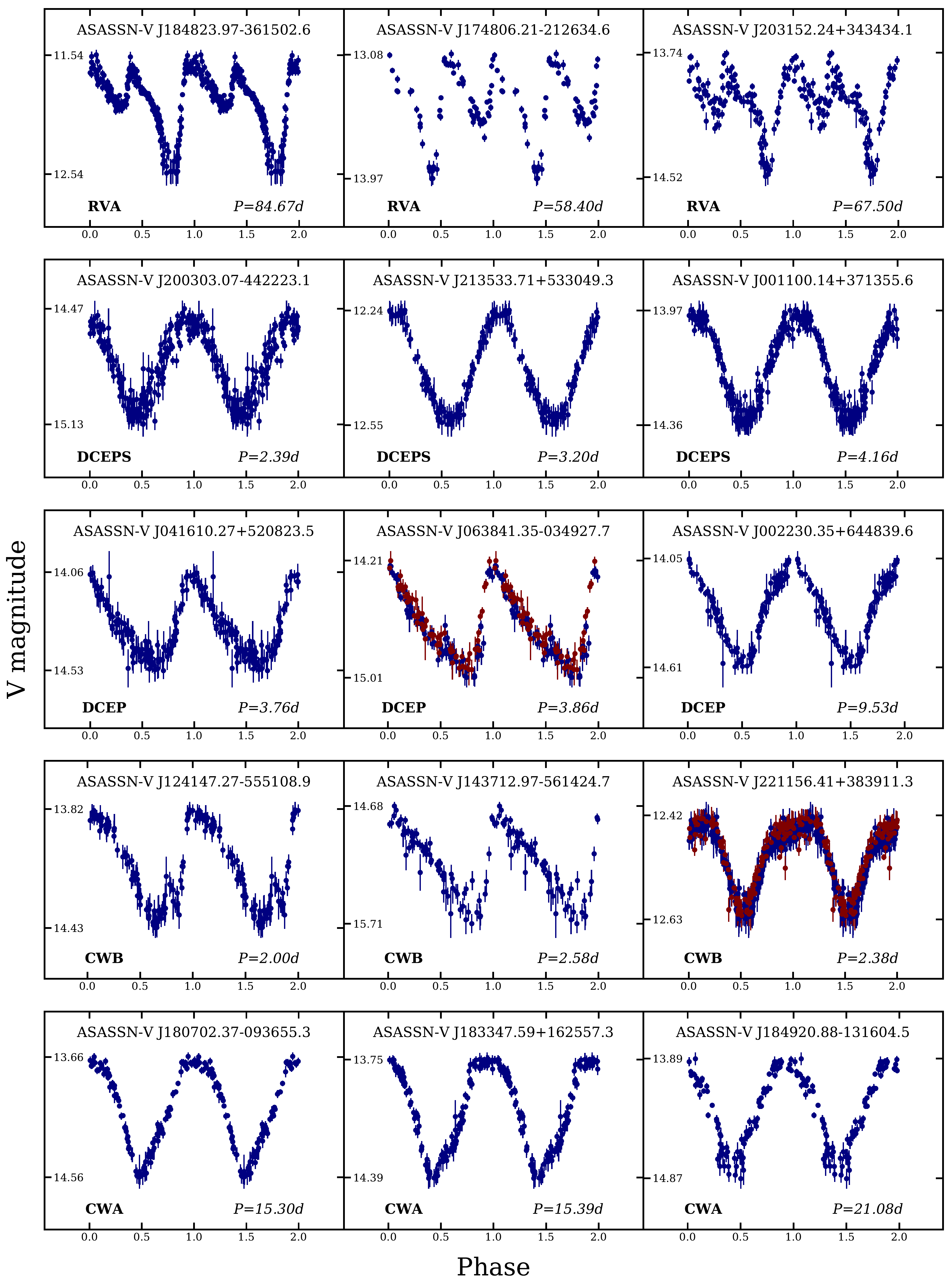}
    \caption{Examples of the DCEP, DCEPS, CWA, CWB and RVA VSX classes. The format is the same as for Fig. \ref{fig:fig9}.}
    \label{fig:fig11}
\end{figure*}

Type II Cepheids overlap classical Cepheids in their periods but are intrinsically less luminous and have distinctive features in their light curves that separate them from classical Cepheids. Type II Cepheids also follow a period-luminosity relation \citep{2006MNRAS.370.1979M}. BL Herculis type variables (CWB) have periods $P<8$ d and have a unique bump along the descending branch in their phased light curves. W Virginis type variables (CWA) have periods $P>8$ d and light curves whose shapes are dependent on the period of pulsation. RV Tauri variables (RVA/RVB) have periods $30<P<150$ d and have alternating primary/secondary minima of different depths. The current sample contains 22 CWA, 44 CWB and 3 RVA variables. We show examples of these variable types in Figure \ref{fig:fig11}.

\subsubsection{Periodic Red Pulsators: Miras and Semi-regular variables}

Mira variables (M) are asymptotic giant branch (AGB) stars with high amplitude (typically defined as $A>2.5$ mag in the V-band) pulsations and periods $P>100$ d. Mira variables can be further classified into sub-types based on the shapes of their light curves \citep{1988A&AS...73..181V,1928HDA.....6...49L}, and have also been found to follow a period-luminosity relationship \citep{2008MNRAS.386..313W}. It is not uncommon for Mira variables to show changes in their period \citep{1999PASP..111...94P}. The current catalog contains 1853 new Mira variables.

Semi-regular variables are typically pulsating red giants with some periodicity in their light curves. These variables usually have irregularities in their light curves that distinguish them from other strictly periodic types, mostly due to multi-periodic behavior \citep{1999A&A...346..542K}. We categorize our list of semi-regular variables based on their period. Short period semi-regular variables (SRS) have periods $P<30$ d while semi-regular (SR) variables have periods $P>30$ d. It is likely that a number of the new semi-regular variables are actually OGLE small amplitude variable red giants (OSARGs; \citealt{2004AcA....54..129S}) that follow a set of period-luminosity relations proposed by \citet{2007AcA....57..201S}. We note that the number of semi-regular variables discovered by ASAS-SN (14,006) is $\sim 30\%$ of the total number of semi-regular variables in the VSX catalog.

\begin{figure*}
	\includegraphics[width=0.99\textwidth]{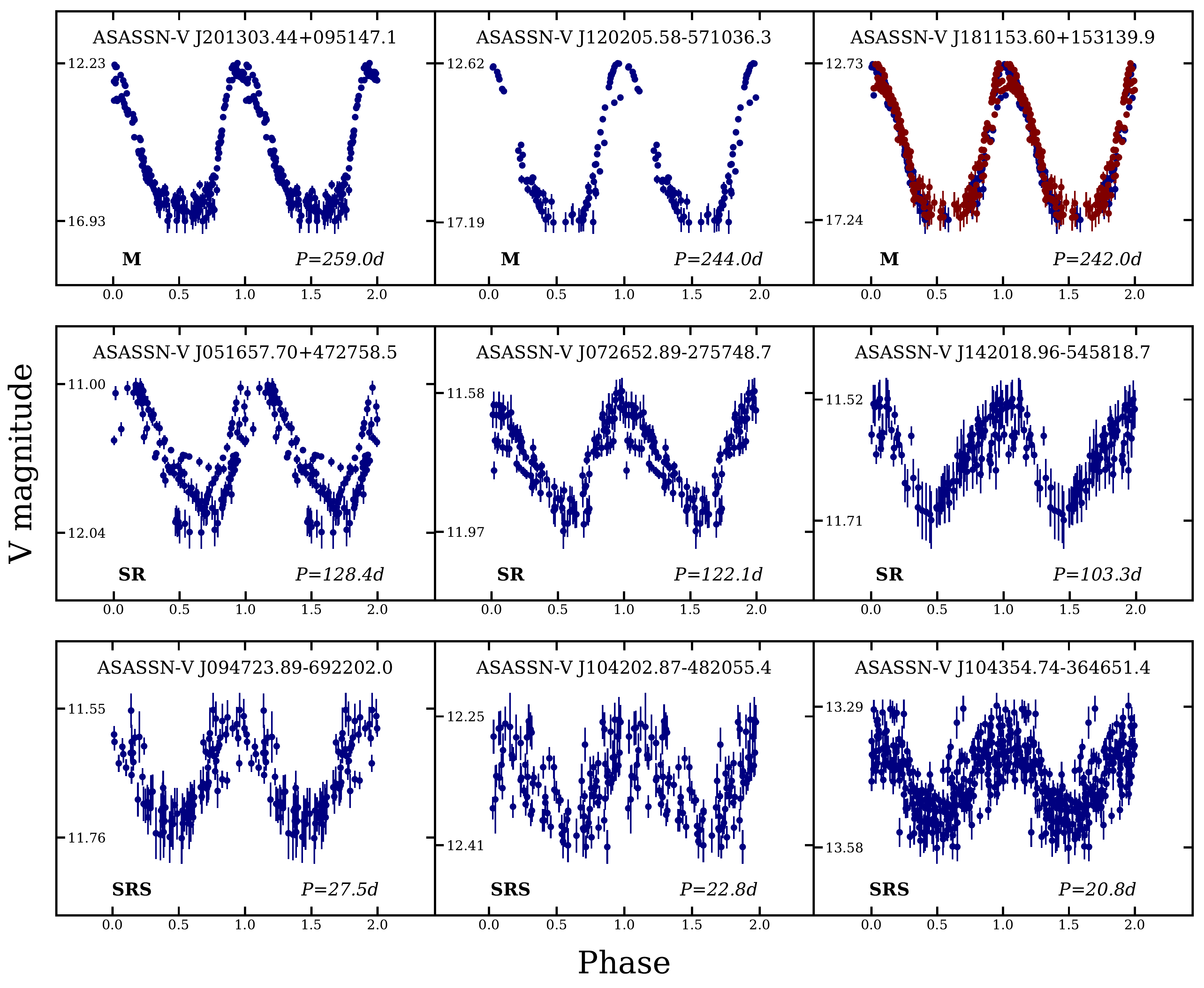}
    \caption{Examples of the periodic red pulsators in our sample, which include the SRS, SR and M VSX classes. The format is the same as for Fig. \ref{fig:fig9}.}
    \label{fig:fig12}
\end{figure*}

Examples of Mira, SRS and SR variables are shown in Figure \ref{fig:fig12}.

\subsubsection{Irregular variables}
\begin{figure*}
	\includegraphics[width=0.99\textwidth]{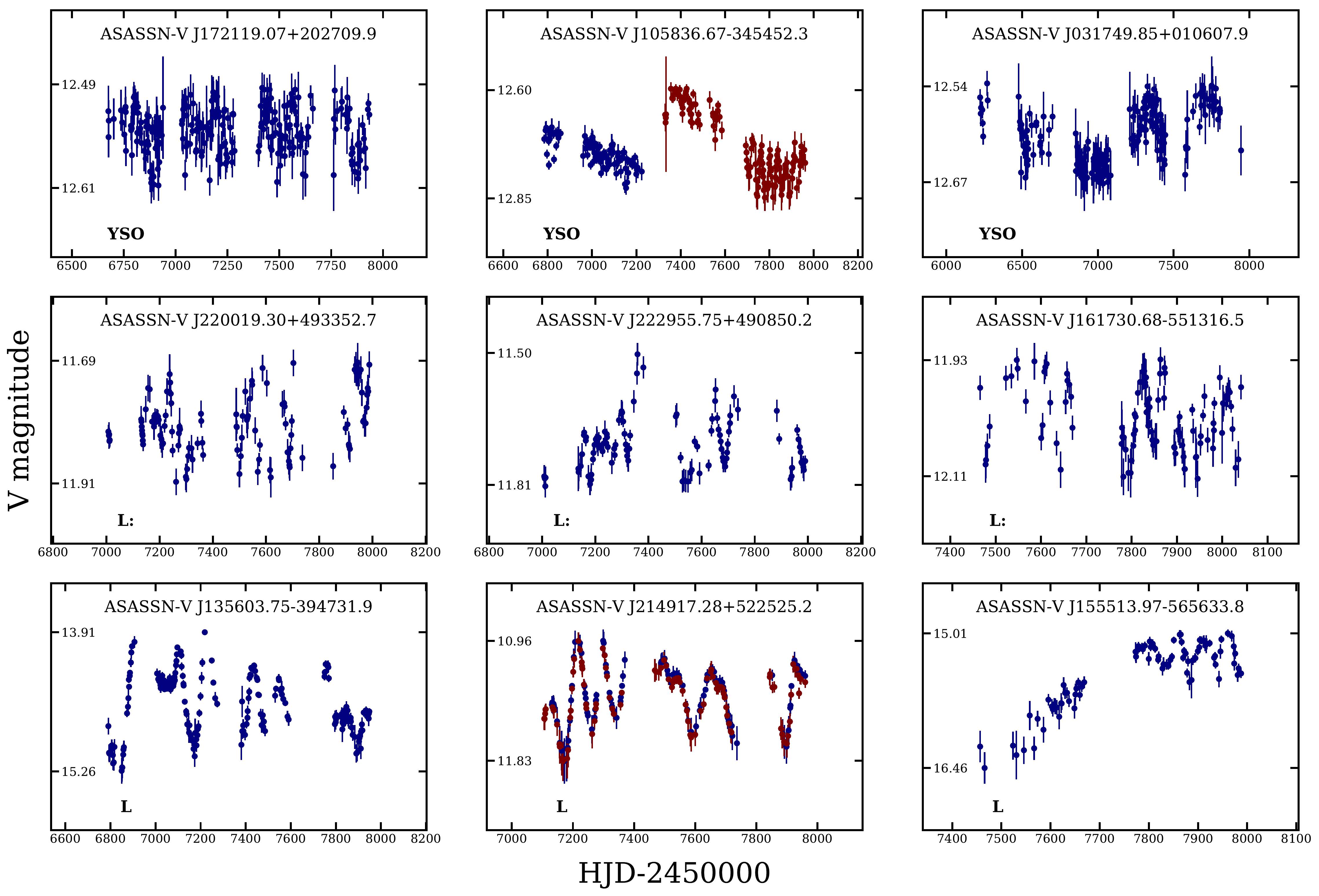}
    \caption{Examples of the irregular variables in our catalog. The format is the same as for Fig. \ref{fig:fig9}.}
    \label{fig:fig13}
\end{figure*}

Our catalog contains several different sub-types of irregulars separated by their colors (see $\S3.3$ and Table \ref{tab:irrclass} for our classification criteria). The majority ($\sim 75\%$, see Table \ref{tab:var}) are red irregulars (L/L:). This catalog increases the number of red irregular variables in the VSX catalog by $70\%$. Irregulars with an infrared excess are classified as young stellar objects (YSO). Examples of irregular variables are shown in Figure \ref{fig:fig13}.

\subsection{Eclipsing Binaries}
Eclipsing binaries provide fundamental physical characteristics for stellar systems (see \citealt{2010A&ARv..18...67T}, and references therein). These systems have been successfully used to derive masses and radii of stars in binary systems to a few percent \citep{2010A&ARv..18...67T}. Recently, eclipsing binaries have also been used to derive extragalactic distances. For example, \citet{2013Natur.495...76P} derived a $\sim2\%$ distance to the Large Magellanic cloud using a set of 8 detached eclipsing binaries.

Photometrically, the light curves of eclipsing binaries are categorized into the GCVS/VSX classes: EW, EB and EA. EW (W UMa) type systems have light curves with minima of similar depths and transition smoothly from the eclipse to the out-of-eclipse phase. EB ($\beta$-Lyrae) type systems also possess light curves that transition smoothly between these states, but have minima of significantly different depths. EA (Algol) type systems may not have a secondary minimum. It is relatively straightforward to pinpoint the exact onset and end of the eclipses in an EA system.

These classes correspond to the contact, semi-detached and detached binary configurations. \citet{2002AcA....52..397P} classified the eclipsing binaries in the ASAS sample based on the likely physical configuration of the binary. While this makes sense probabilistically, the distinction between the different physical classes becomes ambiguous as the amplitude/depth of the eclipses decrease \citep{2006MNRAS.368.1311P}. Classifying eclipsing binaries into photometric classes is more intuitive but has its own problems. For example, it is sometimes difficult to distinguish between the EA and EB classes based on the light curve, so we grouped these sources into a joint class `EA$|$EB' to improve our classifications. Based on our visual inspections, most of the binaries classified as EA$|$EB are actually EB types.

\begin{figure*}
	\includegraphics[width=0.99\textwidth]{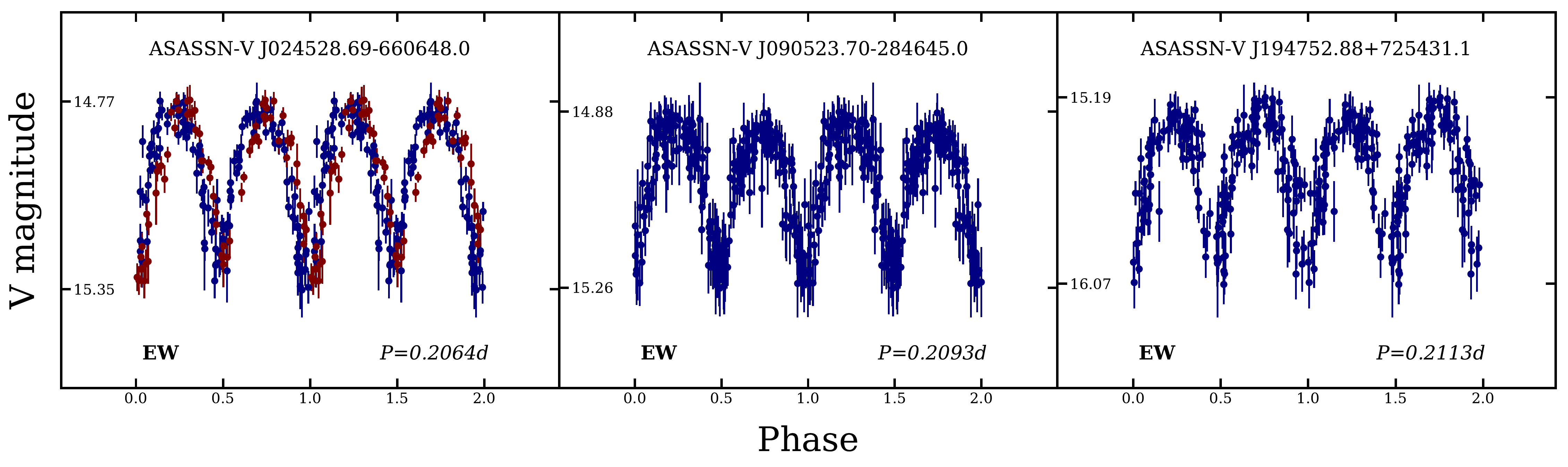}
    \caption{Phased light curves for the three ultra short period contact binaries in the catalog. The format is the same as for Fig. \ref{fig:fig9}.}
    \label{fig:fig14}
\end{figure*}

EW type contact binaries typically have a lower limit to their periods of $P>0.22$ d \citep{1992AJ....103..960R}. We find 3 EW binaries with ultra-short periods, ASASSN$-$V J024528.69$-$660648.0 (P=0.20644d), ASASSN-V J090523.70$-$284645.0 (P=0.20930d), and ASASSN-V J194752.88+725431.1 (P=0.21131d), and these are shown in Figure \ref{fig:fig14}. The VSX catalog lists $>350$ EW binaries with periods $P<0.22$ d, thus it is likely that the lower limit for EW periods are shorter than the limit imposed here.

\begin{figure*}
	\includegraphics[width=0.99\textwidth]{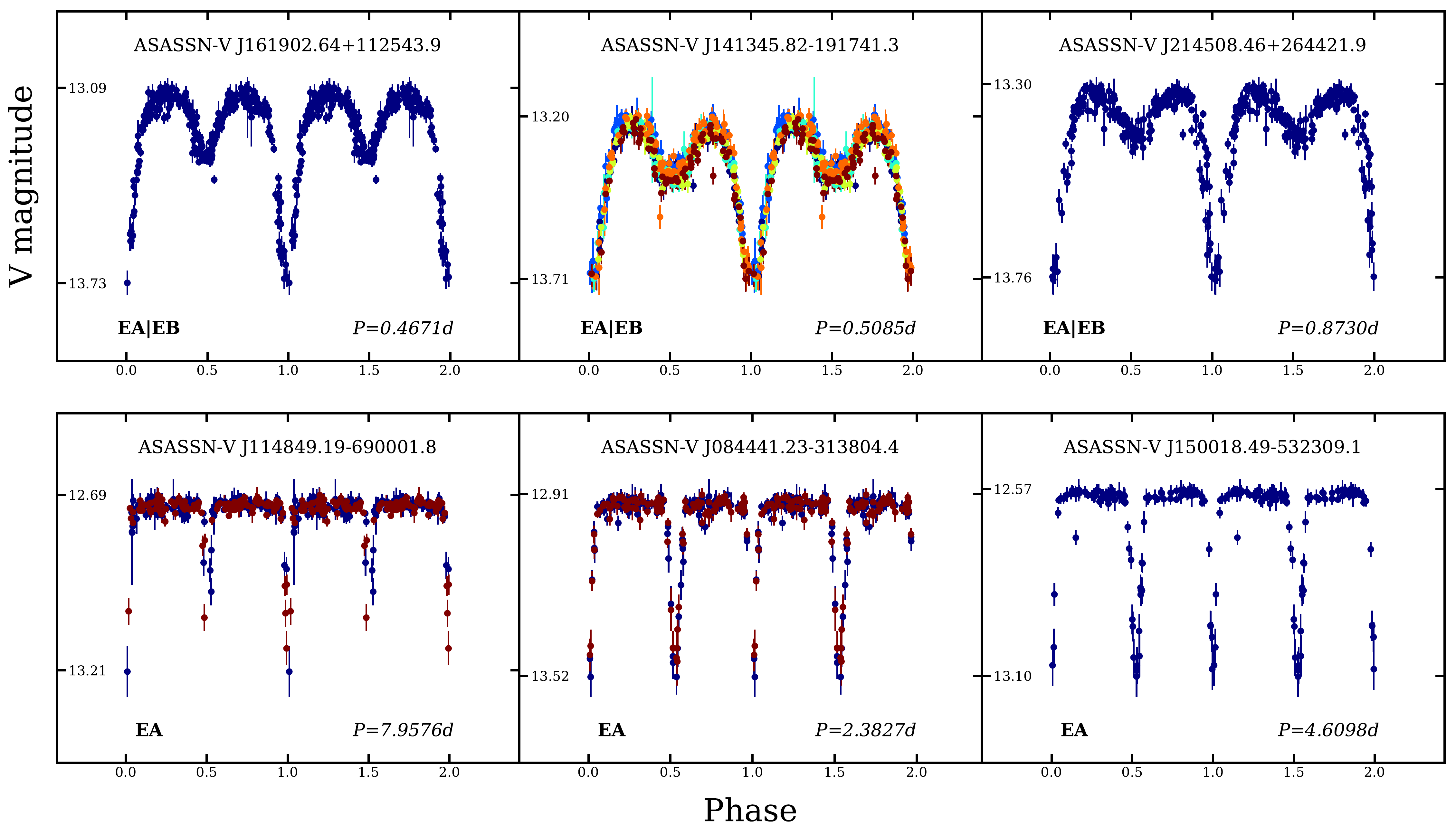}
    \caption{Examples of the EA$|$EB and EA classes. The format is the same as for Fig. \ref{fig:fig9}.}
    \label{fig:fig15}
\end{figure*}

We identify 207 short period ($P<1$ d) detached binaries. Recent evidence suggests that most short period contact binaries are triple systems with tertiary companions (see \citealt{2007AJ....134.2353R}, and references therein). The Lidov-Kozai mechanism \citep{1962P&SS....9..719L,1962AJ.....67..591K} is likely to have created these systems from short period detached binaries \citep{1979A&A....77..145M,1998MNRAS.300..292K,2001ApJ...562.1012E,2006MNRAS.368.1311P,2012JASS...29..145E}. 
We identified 2427 EW binaries, 529 EA|EB binaries and 1530 EA binaries in our data. The eclipsing binaries in our sample span a range of periods from $0.20<P<68.0$ d. Examples of the EA|EB and EA classes are shown in Figure \ref{fig:fig15}.

\subsection{Rotational Variables}

\begin{figure*}
	\includegraphics[width=0.99\textwidth]{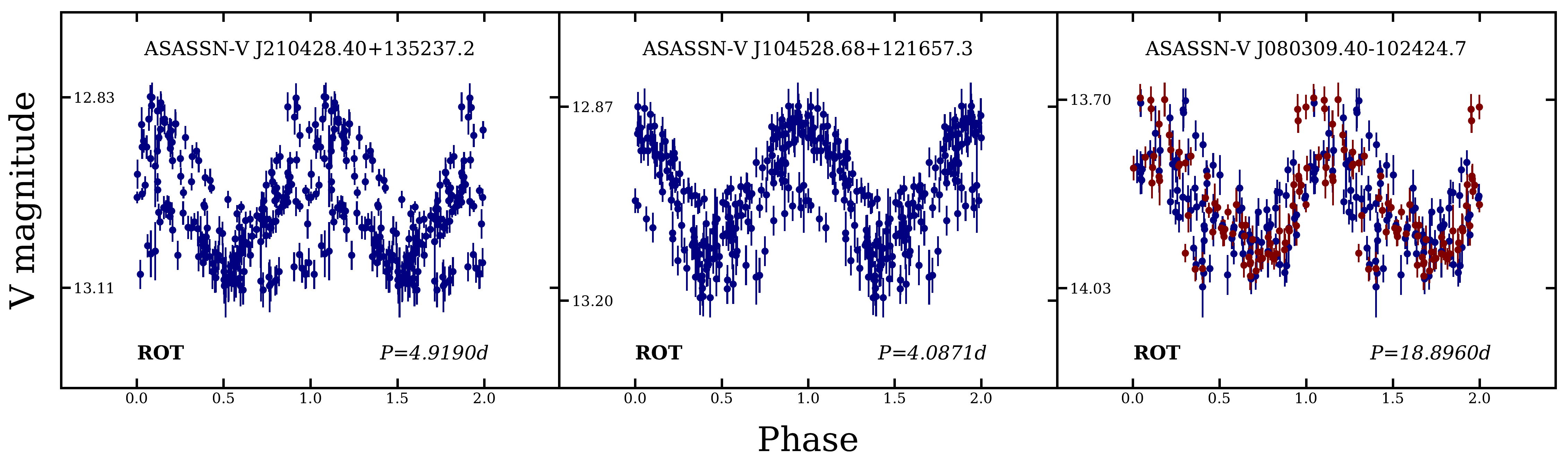}
    \caption{Examples of the rotational variables. The format is the same as for Fig. \ref{fig:fig9}.}
    \label{fig:fig16}
\end{figure*}

Light curves showing evidence of rotational modulation were classified as a generic rotational variable (ROT). We report the best period derived for each light curve and note that this may not be the actual period of rotation for the star. This class is likely to contain numerous RS Canum Venaticorum-type (RS) binary systems along with BY Draconis-type variables and spotted T Tauri stars showing periodic variability. We identified 662 new ROT variables with examples shown in Figure \ref{fig:fig16}.

\citet{2017ATel11034....1J} recently reported the discovery of a likely extreme spotted variable ASASSN-V J211256.97+313724.6. This source had a period of $P=5.412$ d and displayed amplitude modulation from $A=0.58$ mag in 2014 to $A=0.16$ mag in 2014. ASAS data also showed evidence of high amplitude variability, suggesting an extreme degree of star spot coverage, persisting for $\sim 10$ years. Based on these lines of evidence, ASASSN-V J211256.97+313724.6 is most likely a RS binary system. Similar variables are likely to be found in our overall sample of 662 ROT variables.

\subsection{GCAS variables}
The VSX catalog defines GCAS variables to be eruptive irregular variables of early spectral types (O9-A0 III-Ve) with mass outflow from their equatorial zones \citep{2006SASS...25...47W}. A GCAS variable, ASASSN-V J010932.93+614659.0, was identified based on its light curve morphology during the early stages of the variability classification process by \citet{2017ATel10710....1J}. A low resolution spectrum of this object taken with the MDM 2.4m telescope and OSMOS spectrometer showed Balmer and He I lines in absorption, with a strong narrow-H$\alpha$ profile in emission. These features are characteristic of a Be (GCAS) star. Subsequently,  \citet{2017ATel10713....1S} obtained a medium resolution spectrum with the FRODOSpec spectrograph on the Liverpool Telescope, deriving a spectral type of B6V and confirming that this source is a rapid rotator.

\begin{figure*}
	\includegraphics[width=0.99\textwidth]{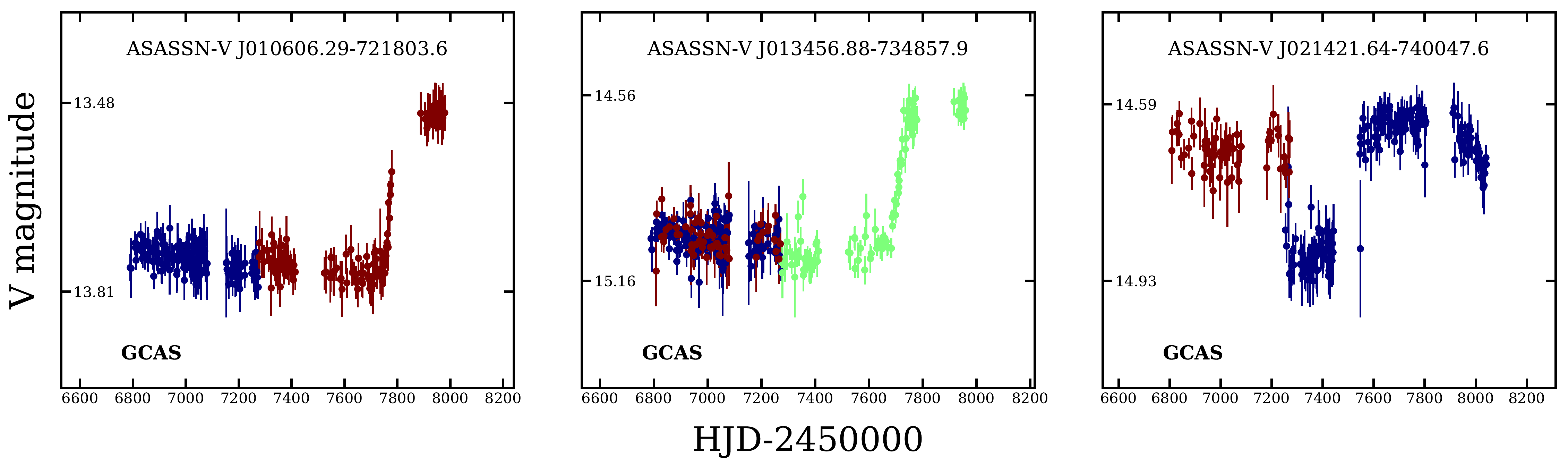}
    \caption{Examples of the GCAS variables. The format is the same as for Fig. \ref{fig:fig9}.}
    \label{fig:fig17}
\end{figure*}

We identified 93 other GCAS variables in our data based on the criteria described in Section $\S3.3$. The locations of these GCAS variables in the 2MASS color-color diagram are highlighted in Figure \ref{fig:fig8}. Each of these sources were visually reviewed to identify outbursts similar to ASASSN-V J010932.93+614659.0 and other Be star outbursts \citep{2018AJ....155...53L}. Based on our selection criteria, it is very likely that most of these sources are rapidly-rotating early-type stars. Examples are shown in Figure \ref{fig:fig17}.

\subsection{Microlensing events}
\begin{figure}
	\includegraphics[width=0.5\textwidth]{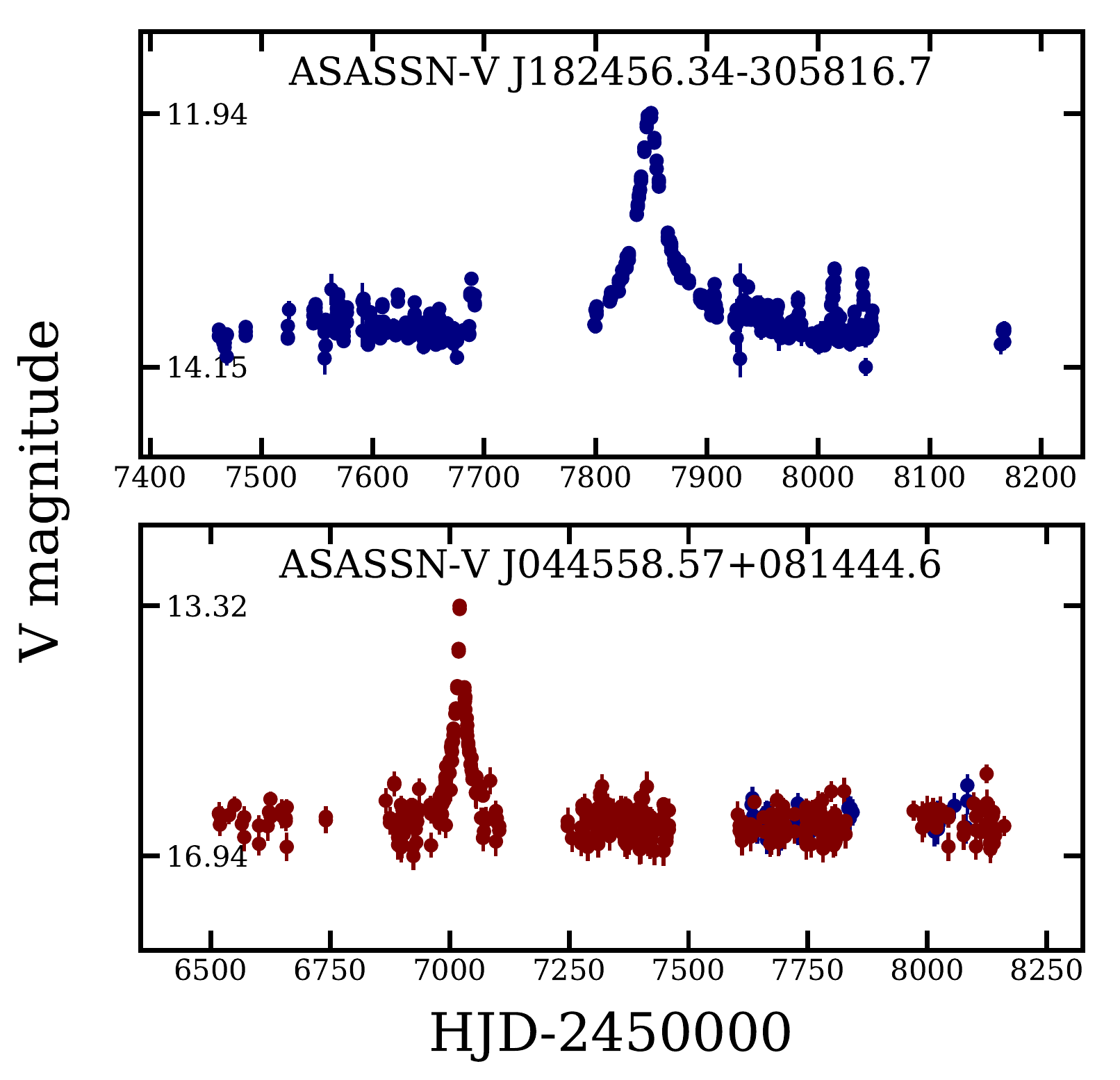}
    \caption{Light curves for the microlensing events ASASSN-V J182456.34$-$305816.7 and ASASSN-V J044558.57+081444.6. The format is the same as for Fig. \ref{fig:fig9}.}
    \label{fig:fig18}
\end{figure}

We discovered two bright, archival microlensing events in our data \citep{2017ATel10677....1J,2017ATel10740....1J} during the variability classification process (Figure \ref{fig:fig18}). ASASSN-V J182456.34$-$305816.7 was discovered near the Galactic bulge with a peak magnification of 16, and ASASSN-V J044558.57+081444.6 was discovered close to the Galactic anti-center with a peak magnification of 33. It is likely that more archival microlensing events are hidden within our data. We are currently conducting a systematic search of the irregular variables to find additional archival microlensing events.

\subsection{Other discoveries}
\begin{figure}
	\includegraphics[width=0.5\textwidth]{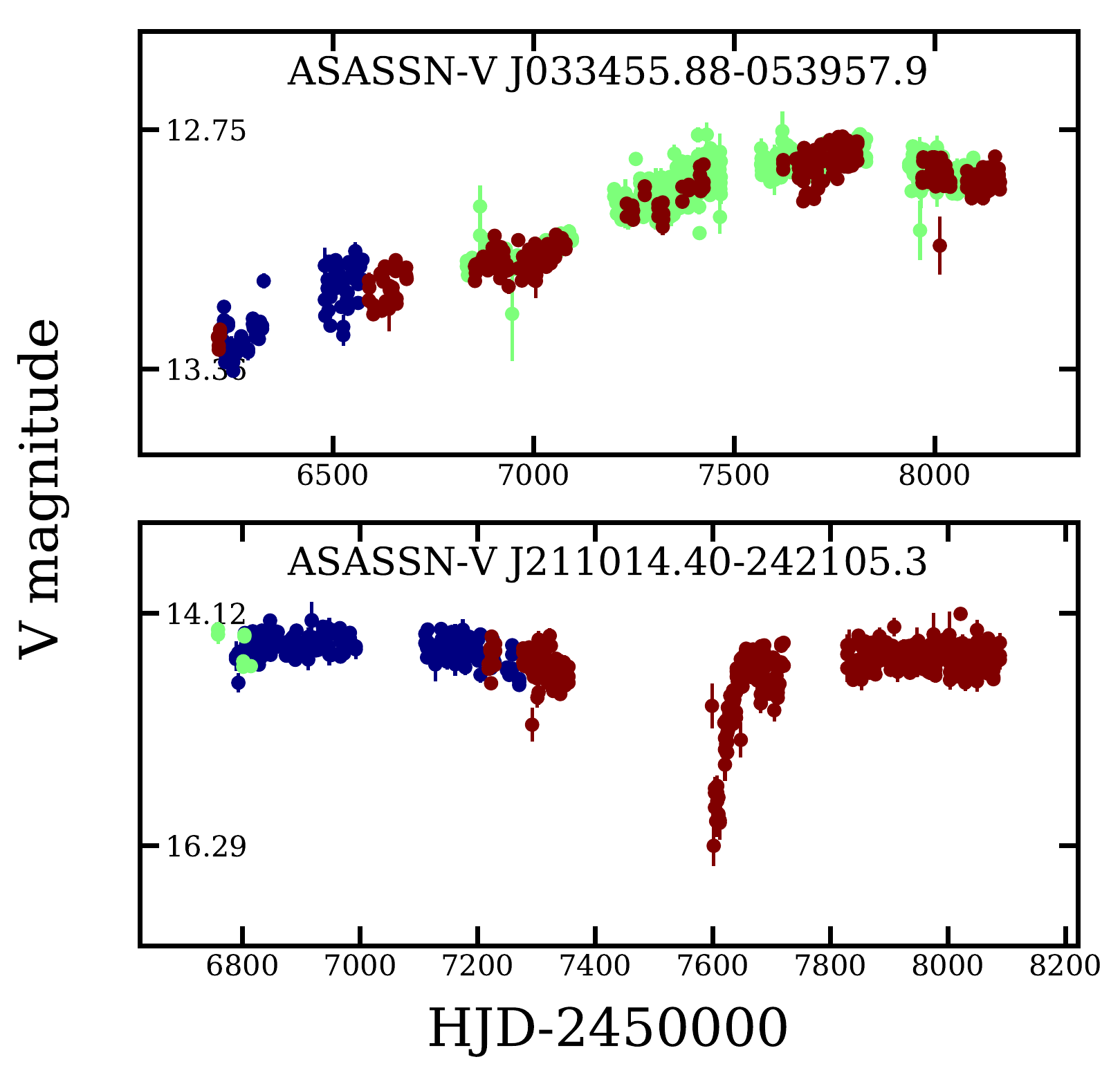}
    \caption{Light curves for the unusual sources ASASSN-V J033455.88$-$053957.9 and ASASSN-V J211014.40$-$242105.3. The format is the same as for Fig. \ref{fig:fig9}.}
    \label{fig:fig19}
\end{figure}

\citet{2017ATel10634....1J} discovered two variables: ASASSN-V J033455.88$-$053957.9 and ASASSN-V J211014.40$-$242105.3, with unusual light curves (Figure \ref{fig:fig19}). ASASSN-V J033455.88$-$053957.9 is a probable early type M-dwarf that brightened by 0.5 mag over a period of $\sim 1,800$ days. It appears in a number of X-ray catalogs, including the Swift X-ray point source catalogue \citep{2014ApJS..210....8E}. The light curve characteristics of this source are atypical of an M-dwarf. 

ASASSN-V J211014.40$-$242105.3 was spectroscopically classified as a cataclysmic variable by \citet{2015MNRAS.453.1879K}. We found a fading event lasting at least 150 days in the ASAS-SN light curve. This fading event is a probable result of episodic mass transfer from the companion star \citep{1999MNRAS.305..225L}, suggesting that ASASSN-V J211014.40$-$242105.3 is likely a cataclysmic variable (CV) of the VY Scl type.

\section{Conclusions}
\label{conclude}
In this paper, we provide a catalog of new variable stars identified by ASAS-SN during our search for supernovae. We classified the variables by:

\begin{enumerate}
  \item Cross-matching our list of sources with existing variable star catalogs to remove known variable sources;
  \item Using an open source random forest classifier \textit{Upsilon} \citep{upsilon} to derive initial classifications for our list of variable sources;
  \item Visually reviewing the \textit{Upsilon} classifications to select a sample of accurately classified variables while flagging irregular variables and removing bad sources;
  \item Training a random forest classifier using the ASAS-SN data and their classifications; and
  \item Reclassifying all the sources along with any new variable sources discovered in the interim.
  
\end{enumerate}

The resulting catalog of 66,179 bright (V$\lesssim17$ mag) new variables is dominated by red pulsators ($\gtrsim 50\%$ of the catalog) and makes significant contributions to the current list of known semi-regular and irregular variables. We also provide lists of common variables (RR Lyrae, eclipsing binaries, Cepheids, Miras, etc.) sampled over the whole sky. While this catalog is not complete for low amplitude variability, it goes a long way towards providing a largely complete sample of high amplitude variables (including Miras and high amplitude semi-regular variables) down to V$\lesssim16$ mag. We estimate the classification errors in this catalog to be $\lesssim 10\%$, with the majority of the errors resulting from confusion between similar variable classes (e.g. sub-classes of eclipsing binaries).

The complete list of ASAS-SN variables along with their V-band light curves are provided online at the ASAS-SN Variable Stars Database (\href{https://asas-sn.osu.edu/variables}{https://asas-sn.osu.edu/variables}). Light curves for bright, known variables will be added to this database over the next year. The next step in our search for variable stars is to extract light curves for all $\gtrsim 50$ million APASS stars to carry out a systematic search.

\section*{Acknowledgements}
We thank the Las Cumbres Observatory and its staff for its
continuing support of the ASAS-SN project. We also thank the Ohio State University College of Arts and Sciences Technology Services for helping us set up the ASAS-SN variable stars database.

ASAS-SN is supported by the Gordon and Betty Moore
Foundation through grant GBMF5490 to the Ohio State
University and NSF grant AST-1515927. Development of
ASAS-SN has been supported by NSF grant AST-0908816,
the Mt. Cuba Astronomical Foundation, the Center for Cos-
mology and AstroParticle Physics at the Ohio State Univer-
sity, the Chinese Academy of Sciences South America Center
for Astronomy (CAS- SACA), the Villum Foundation, and
George Skestos. 

TAT is supported in part by Scialog Scholar grant 24215 from the Research Corporation. Support for JLP is provided in part by FONDECYT through the grant 1151445 and by the Ministry of Economy, Development, and Tourism's Millennium Science Initiative through grant IC120009, awarded to The Millennium Institute of Astrophysics, MAS. SD acknowledges Project 11573003 supported by NSFC. Support for MP has been provided by the PRIMUS/SCI/17 award from Charles University. 

This research was made possible through the use of the AAVSO Photometric All-Sky Survey (APASS), funded by the Robert Martin Ayers Sciences Fund. This publication makes use of data products from the Two Micron All Sky Survey, which is a joint project of the University of Massachusetts and the Infrared Processing and Analysis Center/California Institute of Technology, funded by the National Aeronautics and Space Administration and the National Science Foundation. 

This research has made use of the VizieR catalogue access tool, CDS, Strasbourg, France. The original description of the VizieR service was published in A\&AS 143, 23. 

This research made use of Astropy, a community-developed core Python package for Astronomy (Astropy Collaboration, 2013).





\bibliographystyle{plainnat}

\bsp	
\label{lastpage}
\end{document}